\documentclass[a4paper,usenatbib]{mnras}
\usepackage{aas_macros}
\usepackage{graphicx}
\usepackage{amsmath}
\usepackage{amssymb}
\usepackage[usenames]{xcolor}
\hypersetup{colorlinks=true,citecolor=cyan}

\newcommand{\p}{\ensuremath{\partial}}
\newcommand{\del}{\ensuremath{\delta}}
\newcommand{\Del}{\ensuremath{\Delta}}
\newcommand{\lam}{\ensuremath{\lambda}}

\newcommand{\gam}{\ensuremath{\gamma}}
\newcommand{\Gam}{\ensuremath{\Gamma}}
\newcommand{\sig}{\ensuremath{\sigma}}

\newcommand{\delc}{\ensuremath{\delta_{\rm c}}}
\newcommand{\nuc}{\ensuremath{\nu_{\rm c}}}
\newcommand{\epc}{\ensuremath{\epsilon_{\times}}}
\newcommand{\Sc}{\ensuremath{S_{\times}}}
\newcommand{\So}{\ensuremath{S_{0}}}
\newcommand{\delo}{\ensuremath{\delta_{0}}}

\newcommand{\Msun}{\ensuremath{M_{\odot}}}
\newcommand{\Mh}{\ensuremath{h^{-1}M_{\odot}}}

\newcommand{\avg}[1]{\ensuremath{\left\langle \,#1\, \right\rangle}}

\newcommand{\der}{\ensuremath{{\rm d}}}
\newcommand{\dir}{\ensuremath{\delta_{\rm D}}}
\newcommand{\HT}{\ensuremath{\Theta_{\rm H}}}

\newcommand{\erf}[1]{\ensuremath{{\rm erf}\left(#1\right)}}

\newcommand{\trp}{\ensuremath{{\rm Tr\,}\psi}}

\newcommand{\Xv}{ {\bf X}}

\newcommand{\eqn}[1]{equation~\eqref{#1}}
\newcommand{\eqns}[1]{equations~\eqref{#1}}

\newcommand{\ph}[1]{\phantom{#1}}

\newcommand{\be}{\begin{equation}}
\newcommand{\ee}{\end{equation}}
\newcommand{\Cal}[1]{\ensuremath{\mathcal{#1}}}

\newcommand{\cetal}{CPS16}

\title[Excursion set peaks with shear]
      {Excursion set peaks: the role of shear} 
\date{draft}

\author[E. Castorina, et al.]
{Emanuele Castorina$^{1,2,3}$\thanks{E-mail: ecastorina@berkeley.edu}, Aseem Paranjape$^{4,5}$, Oliver Hahn$^{6,5}$ \& Ravi K. Sheth$^{7,8}$\\  
 $^1$ Berkeley Center for Cosmological Physics, University of California, Berkeley, CA 94720, USA\\
$^2$ Lawrence Berkeley National Laboratory, 1 Cyclotron Road, Berkeley, CA 93720, USA\\
 $^3$ SISSA - International School For Advanced Studies,
      Via Bonomea, 265 34136 Trieste, Italy\\
 $^4$ Inter-University Centre for Astronomy \& Astrophysics,
      Ganeshkhind, Post Bag 4, Pune 411007, India\\
 $^5$ ETH Zurich, Department of Physics, 
      Institute for Astronomy, Wolfgang-Pauli-Strasse 27, 
      CH-8093 Zurich, Switzerland\\
 $^6$ Laboratoire Lagrange, Universit\'e C\^ote d'Azur, Observatoire de la C\^ote d'Azur, CNRS, 
      Blvd de l'Observatoire,\\\hskip0.15in CS 34229, 06304 Nice cedex 4, France\\
 $^7$ Center for Particle Cosmology, University of Pennsylvania, 
      209 S. 33rd St., Philadelphia, PA 19104, USA\\
 $^8$ The Abdus Salam International Center for Theoretical Physics,
      Strada Costiera, 11, Trieste 34151, Italy}
\begin{document}
\label{firstpage}
\pagerange{\pageref{firstpage}--\pageref{lastpage}}

\maketitle 

\begin{abstract}
Recent analytical work on the modelling of dark halo abundances and clustering has demonstrated the advantages of combining the excursion set approach with peaks theory. 
We extend these ideas and introduce a model of excursion set peaks that incorporates the role of initial tidal effects or shear in determining the gravitational collapse of dark haloes. 
The model -- in which the critical density threshold for collapse depends on the tidal influences acting on protohaloes -- is well motivated from ellipsoidal collapse arguments and is also simple enough to be analytically tractable. 
We show that the predictions of this model are in very good agreement with measurements of the halo mass function and traditional scale dependent halo bias in $N$-body simulations across a wide range of masses and redshift. 
The presence of shear in the collapse threshold means that halo bias is naturally predicted to be nonlocal, and that protohalo densities at fixed mass are naturally predicted to have Lognormal-like distributions. 
We present the first direct estimate of Lagrangian nonlocal bias in $N$-body simulations, finding broad agreement with the model prediction. 
Finally, the simplicity of the model (which has essentially a single free parameter) opens the door to building efficient and accurate non-universal fitting functions of halo abundances and bias for use in precision cosmology. 
\end{abstract}


\section{Introduction}
\label{intro}
\noindent
Structure in the Universe builds up hierarchically, starting from small perturbations in the initial matter density and leading up to a complex cosmic web populated by gravitationally collapsed, virialised `haloes'. Since the statistics of the initial field are well-understood (and Gaussian to an excellent approximation), while the structure of the gravitationally evolved field is far richer and more complex, there is considerable theoretical and practical motivation to build models that can accurately identify the \emph{locations} in the initial conditions where collapse is likely to occur at a later time. The statistics of these locations (their number and spatial distribution, for example) could then be mined for cosmologically interesting information.

The excursion set approach provides a useful and flexible framework to do precisely this, by embedding simplified dynamical models of gravitational evolution into the random fluctuations of the initial conditions. In practice, many different effects (both technical and conceptual) need to be accounted for in this approach; e.g., the gravitational evolution of collapsing regions \citep{gg72,ps74}, the correlated nature of the steps of the random walks in the initial field \citep{ph90,bcek91,ms12}, the fact that collapse occurs preferentially at peaks of the density \citep{bbks86,aj90,lp11,ps12b,psd13}, and the stochastic influence of initial tides or shear on the evolution \citep*{ws79,el95,bm96,smt01,st02,ab10,blp14}. These have been studied in various combinations in the literature, but a single model that incorporates all these effects \emph{and} correctly describes numerical measurements is still lacking. In this paper we make some headway into constructing such a model.

In particular, we will motivate and develop a new model of the excursion sets of peaks that explicitly includes the effects of initial tides or shear in determining the collapse of haloes. Our model has a single free parameter, whose value we will adjust to match the low-mass halo mass function in $N$-body simulations at $z=0$. We will then explore the predictions of the model for a number of observables (mass function, halo bias, protohalo overdensity) across wide ranges of mass and redshift. Importantly, we will present results for qualitatively new `nonlocal' bias coefficients that are predicted by our model, comparing these with numerical estimates of the same.

The paper is structured as follows. We start by setting up our notation and describing our numerical simulations in section~\ref{sec:prereq}. We motivate our model in section~\ref{sec:taubarrier} using measurements of the distribution of overdensities of protohalo patches in the initial conditions that eventually become haloes in our simulations. Sections~\ref{sec:mf} and~\ref{sec:bias} compare our model predictions with, respectively, the halo mass function and scale-dependent halo bias measured in the simulations (including the first estimate in the literature of the new nonlocal bias coefficient). For numerical measurements of bias, we employ the model-independent techniques recently proposed by \citet[][hereafter, \cetal.]{cps17} We summarize and discuss possible future extensions of our model in section~\ref{sec:conclude}. Appendix~\ref{app:details} provides technical details of calculations used in the main text, while Appendix~\ref{app:wtdwalks} describes a Monte Carlo algorithm for generating peak weighted random walks, which we use to validate our analytical approximations. We will work exclusively with Gaussian initial conditions in a flat, $\Lambda$-cold dark matter ($\Lambda$CDM) universe.

\section{Analytical and numerical prerequisites}
\label{sec:prereq}
\subsection{Notation}
\label{subsec:notation}
\noindent
The excursion set peak formalism requires us to track the shear tensor $\psi_{ij}$, the derivative of density $\eta_i$ and the shape tensor $\zeta_{ij}$:
\be
\psi_{ij} = \p_i\p_j\phi~~;~~ \eta_i = \p_i\trp ~~;~~ \zeta_{ij} = \p_i\p_j\trp = \nabla^2\psi_{ij}\,,
\label{variables}
\ee
where $i,j=1,2,3$ and $\phi\equiv (3\Omega_{\rm m}H_0^2/2)^{-1}\Phi_{\rm N}$ where $\Phi_{\rm N}$ is the Newtonian potential, so that $\trp = \nabla^2\phi = \del$, the initial matter overdensity field. 
We will denote $\{\psi_1,\psi_2,\psi_3,\psi_4,\psi_5,\psi_6\} \equiv \{\psi_{11},\psi_{22},\psi_{33},\psi_{12},\psi_{13},\psi_{23}\}$, and similarly for $\zeta_{ij}$.
We will need the following spectral integrals:
\begin{align}
\sig_{j\alpha}^2(R) &= \int \der\ln k\,\Del^2(k)\,k^{2j}\,W_\alpha(kR)^2\,,
\label{spectral}\\
\sig_{j\times,\alpha\beta}^2(R,R_0) &= \int \der\ln k\,\Del^2(k)\,k^{2j}\,W_\alpha(kR)\,W_\beta(kR_0)\,,
\label{sigjx}
\end{align}
where $\Del^2(k)\equiv k^3P(k)/2\pi^2$ is the dimensionless matter power spectrum and where $\alpha,\beta$ can be one of \{T,G\} for TopHat and Gaussian, respectively: $W_{\rm T}(q) \equiv (3/q^3)(\sin q-q\cos q)$ and $W_{\rm G}(q) \equiv {\rm e}^{-q^2/2}$. We will suppress the $R$-dependence when no confusion can arise. 
The spectral ratio \gam\ defined by
\be
\gam\equiv\sig_{\rm 1m}^2/(\sig_{\rm 0T}\sig_{\rm 2G})\,,
\label{gam-def}
\ee
where $\sig_{1\textrm{m}}^2(R)\equiv\sig_{1\times,\textrm{GT}}^2(R,R)$, plays a key role in the cross-correlation between the shape and shear tensors\footnote{The reason for allowing mixed filtering of this type is the fact that we will typically be interested in TopHat filtered density (and shear) fields, while the shape tensor requires Gaussian filtering as discussed by \citet{bbks86}. Although this means that $\zeta_{ij,\textrm{G}}\neq\nabla^2\psi_{ij,\textrm{T}}$, in practice a judicious choice of matching the filtering scales means that this does not introduce any problems. Unless stated otherwise, we will ensure that Gaussian filtered fields use a smoothing scale $R_{\rm G}$ related to the corresponding TopHat filter scale by demanding $\avg{\del_{\rm G}|\del_{\rm T}}=\del_{\rm T}$, i.e. $\avg{\del_{\rm G}\del_{\rm T}}=\sig_{\rm 0T}^2$, which leads to $R_{\rm G}\approx R_{\rm T}/\sqrt{5}$ with a slow variation \citep{psd13}.}.

The following linear combinations of the matrix elements will be useful
\begin{align}
\sig_{\rm 2G}\,x &\equiv -(\zeta_1+\zeta_2+\zeta_3)\notag\\ 
\sig_{\rm 2G}\,y &\equiv -(\zeta_1-\zeta_3)/2\label{xyz-def}\\ 
\sig_{\rm 2G}\,z &\equiv -(\zeta_1-2\zeta_2+\zeta_3)/2\notag\\
&\notag\\
\sig_{\rm 0T}\,\nu &\equiv \psi_1+\psi_2+\psi_3\notag\\ 
\sig_{\rm 0T}\,l_2 &\equiv \sqrt{15}\,(\psi_1-\psi_3)/2 \label{nul2l3-def}\\
\sig_{\rm 0T}\,l_3 &\equiv \sqrt{5}\,(\psi_1-2\psi_2+\psi_3)/2\notag\\
&\notag\\
\sig_{\rm 0T}\,l_A &\equiv \sqrt{15}\,\psi_A\,,\,A=4,5,6\,.\label{lA-def}
\end{align}
The covariance structure of these variables is well known \citep[][see also Appendix~\ref{app:corrs}]{bbks86,vdwb96,lw10}. In particular, the variables $\{\nu,l_2,\ldots,l_6\}$ have zero mean and unit variance by construction, and upon integrating over the shape tensor at random locations they form a set of $6$ independent standard normal variates \citep{vdwb96,smt01}. We will work exclusively in the eigenbasis of the shape tensor $\zeta_{ij}$, so that the variables $\{x,y,z\}$ are linear combinations of the eigenvalues of $\zeta_{ij}$ and the off-diagonal components $\zeta_A$ vanish for $A=4,5,6$. 

We will also frequently use the variable \nuc\ defined as
\be
\nuc(m,z) \equiv \frac{\delc(z)}{\sig_{\rm 0T}(m)}\,\frac{D(0)}{D(z)}\,,
\label{nuc-def}
\ee
where $m=(4\pi/3)R^3\bar\rho$ with $\bar\rho$ the mean density of the Universe, $D(z)$ is the growth factor of linear density perturbations, and $\delc(z)$ is the critical threshold for collapse at redshift $z$ in the spherical collapse model\footnote{The value of $\delc(z)$ in a flat $\Lambda$CDM universe is weakly dependent on redshift and cosmology, in contrast to that in an Einstein-deSitter background \citep[see, e.g.,][]{ecf96}, and can be approximated by $\delc(z) = \del_{\rm c,EdS}(1-0.0123\log_{10}(1+x^3))$, where $x\equiv(\Omega_{\rm m}^{-1}-1)^{1/3}/(1+z)$ and $\del_{\rm c,EdS}=1.686$ \citep{Henry2000}. For example, requiring collapse at present epoch gives $\delc(z=0)=1.676$ for the Planck13 cosmology and $1.674$ for the WMAP3 cosmology.}.

\subsection{$N$-body Simulations}
\noindent
We will validate our model using $N$-body simulations of collisionless CDM in periodic cubic boxes performed using the tree-PM code\footnote{http://www.mpa-garching.mpg.de/gadget/} \textsc{Gadget-2} \citep{springel:2005}. 
For low redshift results ($0\lesssim z \lesssim 1$) we have used two configurations, a large box of comoving size $(2h^{-1}{\rm Gpc})^3$ which samples the high-mass end of the mass function, and a smaller box of size  $(200h^{-1}{\rm Mpc})^3$ to sample lower masses. Additionally, for higher redshifts ($6\lesssim z\lesssim 9$) we have used a box of size $(50h^{-1}{\rm Mpc})^3$. Table~\ref{tab:sims} summarizes the properties of these simulations.

\begin{table}
\centering
\begin{tabular}{cccccc}
\hline
\hline
Cosmo. & $L_{\rm box}$ & $N_{\rm p}$ & $m_{\rm p}$ & $\epsilon$ & $N_{\rm r}$\\
 & $({\rm Mpc}/h)$ &  & $(\Msun/h)$ & $({\rm kpc}/h)$  & \\
\hline
Planck13 & $2000$ & $1024^3$ & $6.5\times10^{11}$ & $65$ & $9$ \\
WMAP3 & $200$ & $512^3$ & $4.1\times10^{9}$ & $12.5$ & $10$ \\
Planck13 & $50$ & $1024^3$ & $1.0\times10^{7}$ & $2$ & $5$ \\
\hline
\hline
\end{tabular}
\caption{Details of $N$-body simulations used in this work. Columns correspond to the cosmology (see text), simulation box length $L_{\rm box}$, number of particles $N_{\rm p}$, particle mass $m_{\rm p}$, force resolution $\epsilon$ (comoving) and number of realisations $N_{\rm r}$. The simulations used a PM grid of $8\times N_{\rm p}$ cells in each case.}
\label{tab:sims}
\end{table}

Initial conditions were generated in each case at $z=99$ employing $2^{\rm nd}$-order Lagrangian Perturbation Theory \citep{scoccimarro98}, using the code\footnote{http://www.phys.ethz.ch/$\sim$hahn/MUSIC/} \textsc{Music} \citep{hahn11-music}. All simulations used flat $\Lambda$CDM cosmologies with transfer functions generated using the prescription by \citet{eh98}.
The low redshift large box and the high redshift box used a `Planck13' cosmology having parameters\footnote{$\Omega_{\rm m}$ and $\Omega_{\rm b}$ are the present-epoch fractional densities of all matter and baryons, respectively, the Hubble constant is $H_0=100h\,$km/s/Mpc, $\sig_8$ is the linear theory r.m.s. matter fluctuation at $z=0$ smoothed in spheres of $8h^{-1}$Mpc and $n_{\rm s}$ is the scalar spectral index.} $\left(\Omega_{\rm m},\Omega_{\rm b},h,\sig_8,n_{\rm s}\right)=\left(0.315,0.0487,0.673,0.83,0.96\right)$ consistent with results presented by the \cite{Planck13-XVI-cosmoparam}, while the low redshift smaller box used a `WMAP3' cosmology with parameters $\left(\Omega_{\rm m},\Omega_{\rm b},h,\sig_8,n_{\rm s}\right)=\left(0.25,0.045,0.7,0.8,0.96\right)$ consistent with the 3-year results of the Wilkinson Microwave Anisotropy Probe \citep{spergel+07}.
We ran multiple realisations for each configuration by changing the random number seeds for the initial conditions.
All simulations were run on the Brutus cluster\footnote{http://www.cluster.ethz.ch/index\_EN} at ETH Z\"urich.

\begin{figure}
\centering
\includegraphics[width=0.45\textwidth]{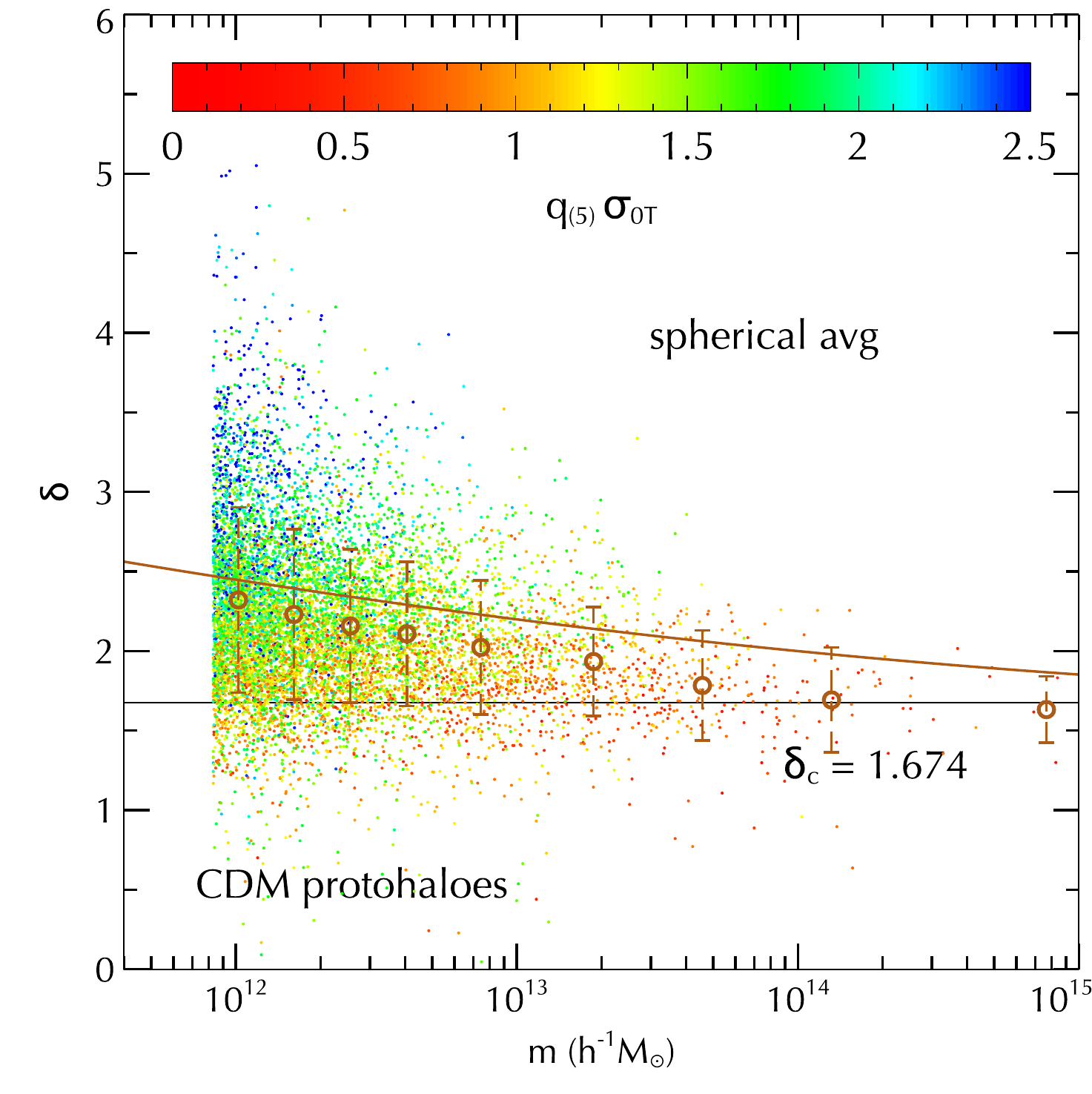}
\caption{Distribution of the $z=0$ protohalo overdensity in the WMAP3 simulation as a function of mass, coloured by the measured values of $q_{(5)}\sig_{\rm 0T}$ (equation~\ref{q5-def}). Horizontal line shows the spherical collapse value $\delc=1.674$. Results are shown for $10^4$ randomly chosen haloes resolved with $200$ particles or more. Empty circles show the median overdensity in bins of halo mass, with error bars indicating the standard deviation in the bin (the error on the mean is typically much smaller). The smooth curve shows the mean barrier $\avg{B|m}$ for the model discussed in section~\ref{sec:mf} which gives a good description of the halo mass function.}
\label{fig:Bq5}
\end{figure}

Haloes were identified using the code\footnote{http://code.google.com/p/rockstar/} \textsc{Rockstar} \citep{behroozi13-rockstar}, which is based on an adaptive hierarchical Friends-of-Friends algorithm in $6$-dimensional phase space. \textsc{Rockstar} has been shown to be robust for a variety of diagnostics such as density profiles, velocity dispersions, merger histories and the halo mass function. Throughout, we will use the mass definition $m_{\rm 200b}$ which is the mass contained in a spherical volume of radius $R_{\rm 200b}$ at which the enclosed density surrounding the center-of-mass reaches $200$ times the mean density of the Universe. Unless specified, results are shown for haloes resolved with $100$ particles or more.

For our analysis, we require measurements of $\psi_{ij}$, $\eta_i$ and $\zeta_{ij}$ at the locations of protohaloes in the initial conditions of our simulations (i.e., patches in the initial conditions which eventually form haloes at, say, $z=0$). We do this following the methodology of \cite{hp14}. Briefly, we use the code \textsc{Music} to generate the initial density used for the simulation as grid data, Fourier transforming which allows us to calculate $\psi_{ij}$, $\eta_i$ and $\zeta_{ij}$ on the grid. Using particle IDs of the consituents of a protohalo patch, we can then directly evaluate these tensors at the patch locations, averaging them component-wise in spheres corresponding to the Lagrangian radius $R_{\rm L} = (3m/4\pi\bar\rho)^{1/3}$ of each patch. Quantities such as eigenvalues and rotational invariants are then evaluated using these averaged tensors. We refer the reader to section 3 of \cite{hp14} for further details.

\section{A simple model for the role of shear}
\label{sec:taubarrier}
\noindent
From the point of view of ellipsoidal collapse, we are interested in a barrier that depends on the rotational invariant $q_{(5)}$ defined as
\be
q_{(5)}^2 \equiv (l_2^2+l_3^2+l_4^2+l_5^2+l_6^2)/5\,,
\label{q5-def}
\ee
whose distribution for random locations is closely linked to a Chi-squared with $5$ degrees of freedom. \citep[See, e.g.,][who discussed an excursion set model using the variable $r^2=q_{(5)}^2\sig_{\rm 0T}^2$.]{st02} A dependence of the barrier on the initial tidal field might be anticipated by arguing that smaller protohalo patches, which do not dominate their surroundings, must necessarily struggle more in order to hold themselves together and become haloes, while larger patches are less susceptible to the disrupting influence of gravitational shear \citep{smt01}. Therefore, the barrier for smaller patches should, on average, be higher than that for larger patches, with a height that grows in proportion to the average tidal field surrounding the patch \citep[see, however][for an alternate point of view, which we return to in section~\ref{sec:conclude}]{blp14}.

\begin{figure}
\centering
\includegraphics[width=0.45\textwidth]{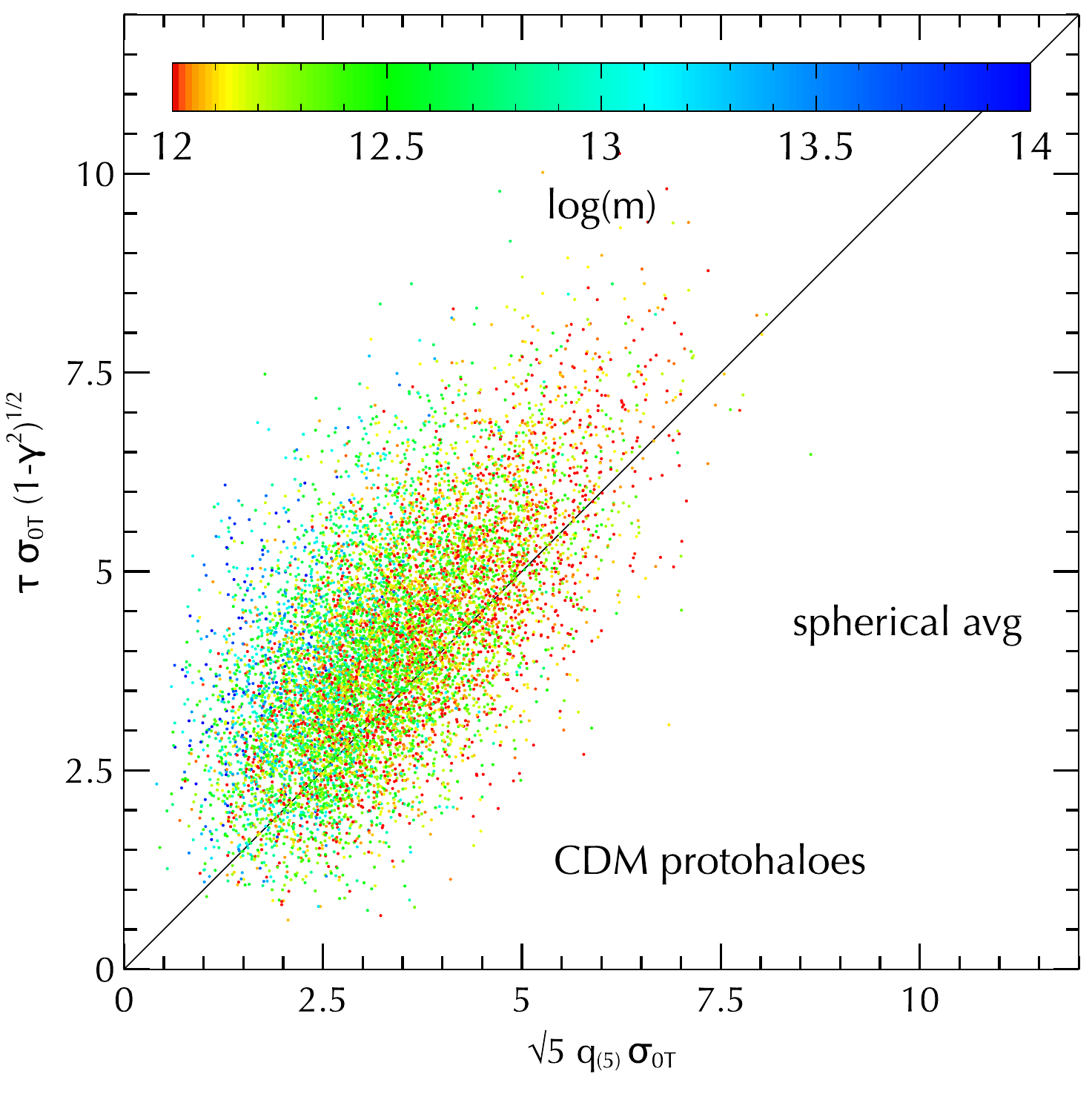}
\caption{Joint distribution of $\tau\sig_{\rm 0T}\sqrt{1-\gam^2}$ (equation~\ref{tau-def}) and $\sqrt{5}q_{(5)}\sig_{\rm 0T}$ for the $z=0$ CDM protohaloes in the WMAP3 simulation. Solid line shows the one-to-one relation. Results are shown for the same haloes as in Figure~\ref{fig:Bq5}.}
\label{fig:tauVsq5}
\end{figure}

We would then like to track the variables $\{\nu,q_{(5)},x,y,z\}$, the latter three being necessary to impose the peaks constraint as in \citet{bbks86}. Figure~\ref{fig:Bq5} shows the correlation between $q_{(5)}\sig_{\rm 0T}$ and the overdensity \del\ of the CDM protohaloes at $z=0$ in the WMAP3 simulation \citep[see also][]{dwbs08,hp14,lbp14}. From the colour scheme we see that $q_{(5)}\sig_{\rm 0T}$ correlates strongly with \del, in stark contrast to a Gaussian random field for which these variables are \emph{independent} \citep{smt01,st02}. The empty circles show the mean overdensity in bins of halo mass, and the error bars show the scatter of the overdensity in each bin. The smooth curve shows the mean excursion set barrier from the model discussed below, in which the barrier slope is adjusted to give a good description of the low-redshift halo mass function in this simulation. We see that there is reasonable agreement between the model barrier and the measured mean overdensities, although the model is systematically higher than the measurements, with the relative gap increasing at high masses. We will discuss this later. 

An excursion set barrier based on $q_{(5)}$ is, however, difficult to analyse due to the correlation between $\{l_2,l_3\}$ and $\{y,z\}$. Instead, as we show in Appendix~\ref{app:mf}, there is a combination of variables that is straightforward to handle as well as being close to $q_{(5)}$ in its statistics. We define the variable $\tau$ as follows:
\begin{align}
\tau^2 &\equiv \frac{3q_{(3)}^2 + \left(l_2-\gam\sqrt{15}\,y\right)^2 + \left(l_3-\gam\sqrt{5}\,z\right)^2}{(1-\gam^2)}\,,
\label{tau-def}
\end{align}
where the quantity $q_{(3)}$ defined by
\be
q_{(3)}^2 \equiv (l_4^2+l_5^2+l_6^2)/3
\label{q3-def}
\ee
captures the misalignment between the shape and shear tensors as represented by the off-diagonal shear components in the shape eigenbasis (see equation~\ref{lA-def}).

In the limit $\gam\to0$ we would have $\tau\to \sqrt{5}\,q_{(5)}$. For non-zero \gam\ it is clear that $\tau$ and $q_{(5)}$ will be tightly correlated for a Gaussian random field. As Figure~\ref{fig:tauVsq5} shows, this is also the case for protohaloes, reassuring us that the combination $\tau\sqrt{(1-\gam^2)/5}$ should indeed be a good proxy for $q_{(5)}$ in the ESP calculation.

\begin{figure}
\centering
\includegraphics[width=0.45\textwidth]{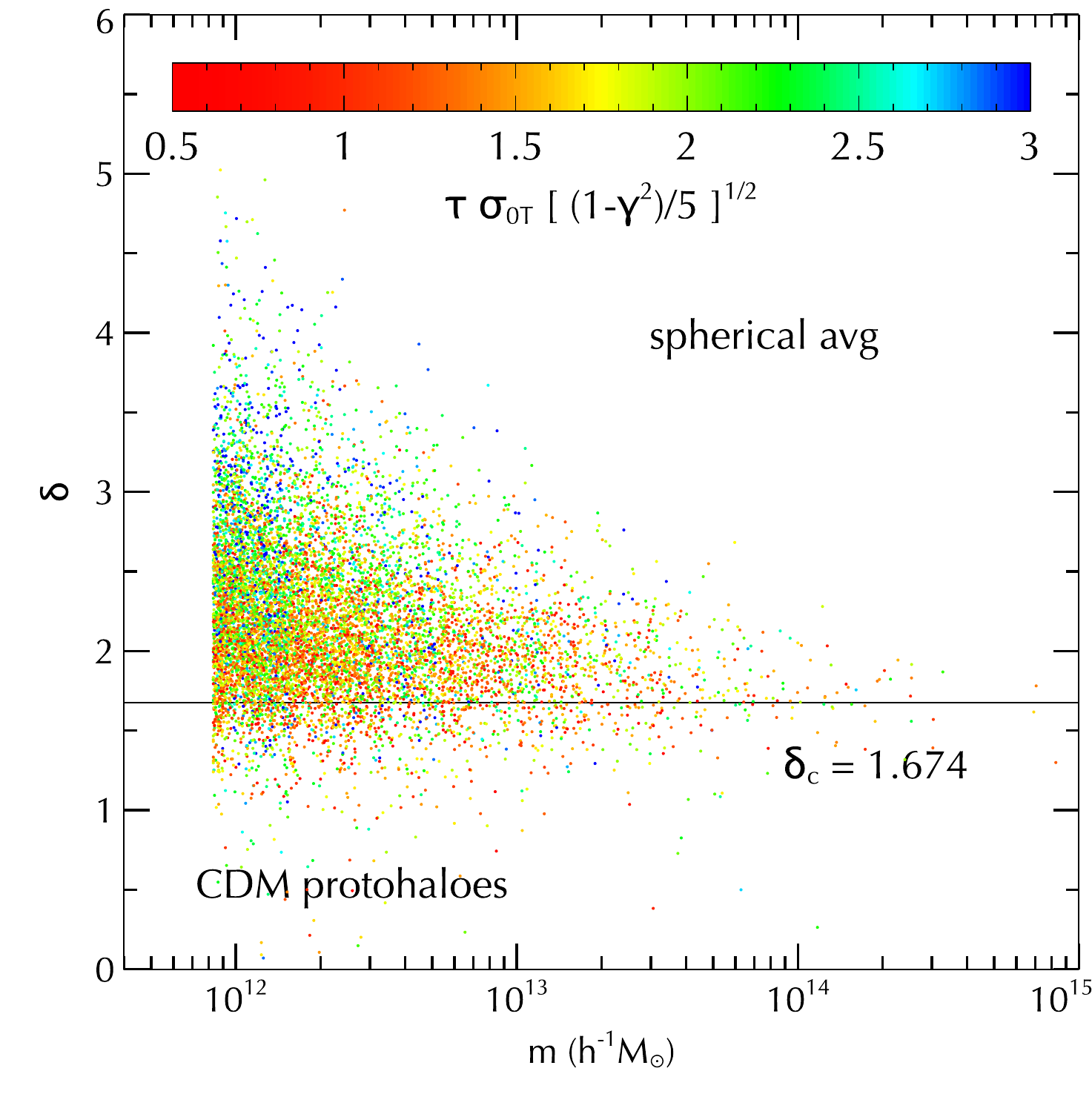}
\caption{Distribution of the $z=0$ protohalo overdensity in the WMAP3 simulation as a function of mass, coloured by the measured values of $\tau\sig_{\rm 0T}\sqrt{(1-\gam^2)/5}$, which is our proxy for $q_{(5)}\sig_{\rm 0T}$. Horizontal line shows the spherical collapse value $\delc=1.674$. Results are shown for the same haloes as in Figure~\ref{fig:Bq5}.}
\label{fig:Btau}
\end{figure}

\begin{figure*}
\centering
\includegraphics[width=0.325\textwidth]{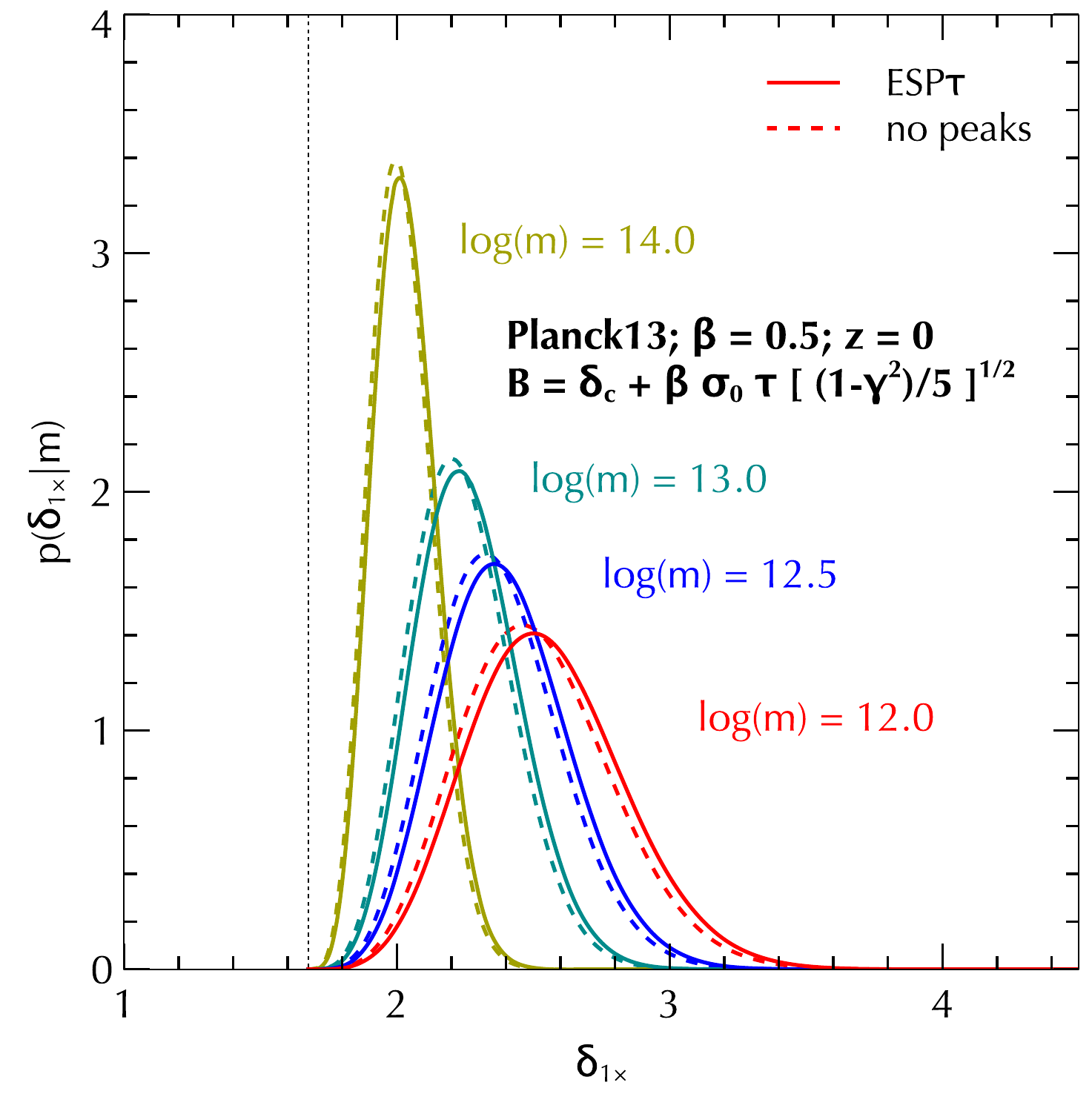}
\includegraphics[width=0.325\textwidth]{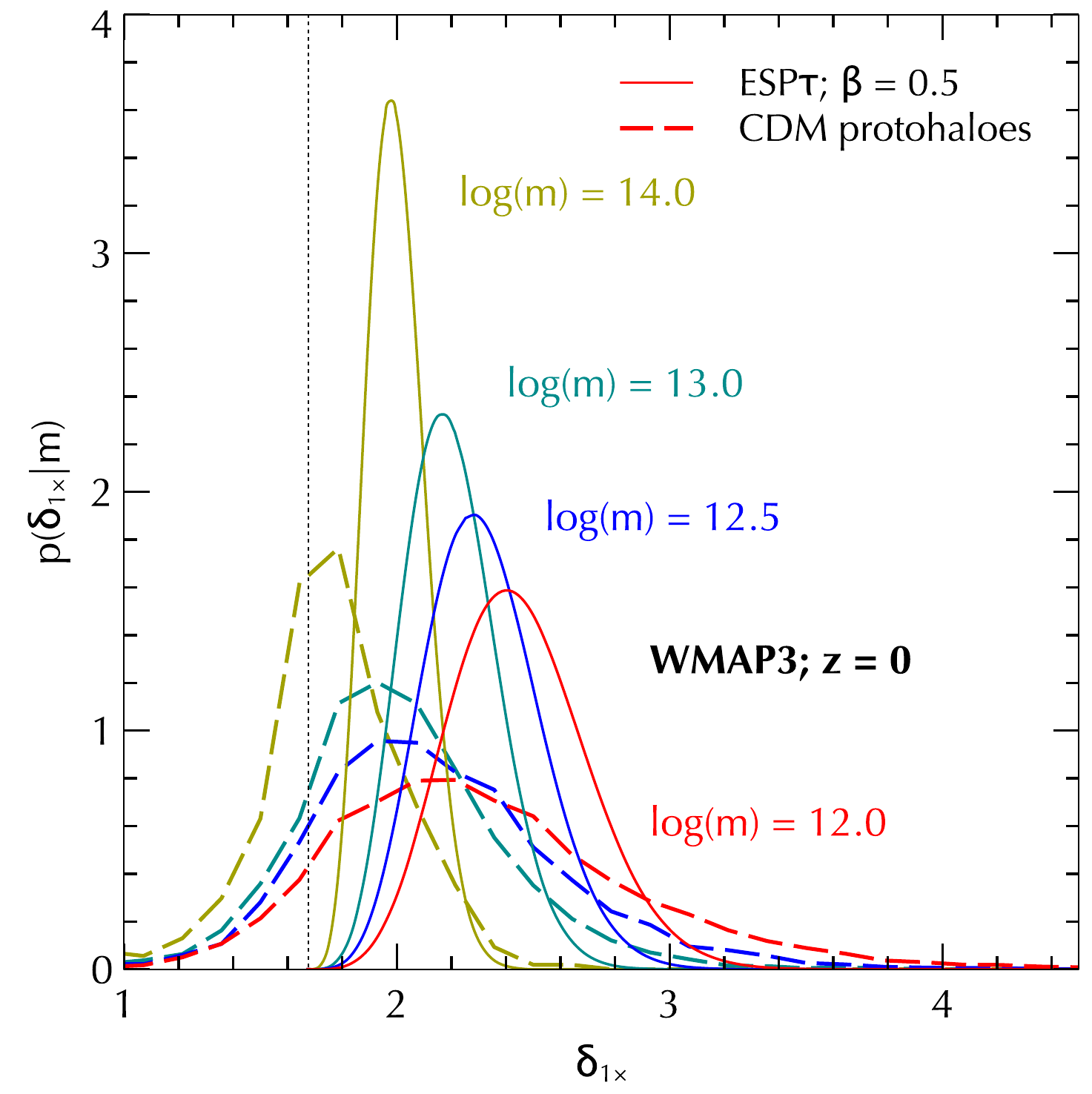}
\includegraphics[width=0.325\textwidth]{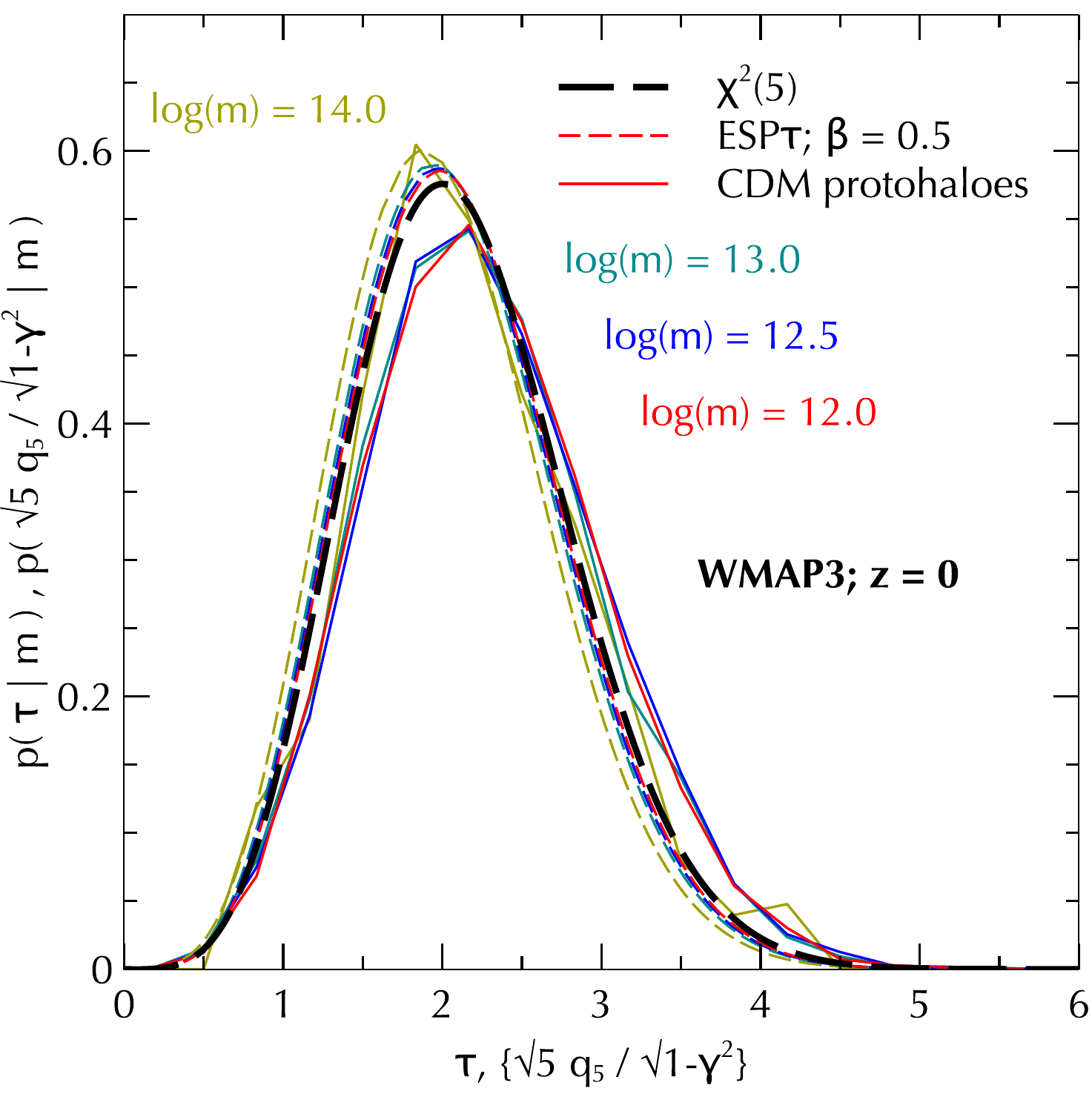}
\caption{\emph{(Left panel:)} Predicted distribution of proto-halo density $\delta_{1\times}$ in excursion set models using the barrier \eqref{tau-barrier} with (solid) and without (dashed) the peaks constraint, for the Planck13 cosmology. See text for details. \emph{(Middle panel:)} Comparison of the distribution of $\delta_{1\times}$ predicted by the ESP$\tau$ model (solid lines) with corresponding measurements using CDM protohaloes in the WMAP3 simulation (dashed lines). \emph{(Right panel:)} Comparison of the distribution $p(\tau|m)$ predicted by the ESP$\tau$ model (thin dashed lines) with the measured distribution of $\sqrt{5}q_{(5)}/\sqrt{1-\gam^2}$ using the WMAP3 CDM protohaloes (solid lines). The thick dashed line marked $\chi^2(5)$ shows the mass-independent distribution $p_5(\tau)$ which is closely linked to the Chi-squared distribution with 5 degrees of freedom.}
\label{fig:Bstoch}
\end{figure*}

The distribution of $\tau$ in the eigenbasis of $\zeta_{ij}$ is given by (Appendix~\ref{app:mf})
\be
\der\tau \, p_5(\tau) \equiv \frac{\der\tau}{3}\,\sqrt{\frac2\pi}\,\tau^4\,{\rm e}^{-\tau^2/2}\,,
\label{p5tau-def}
\ee
which is related to the distribution of a Chi-squared variate with 5 degrees of freedom: $\der \tau\,p_5(\tau) = \der \tau^2\,p_{\chi^2}(\tau^2;k=5)$. 
A virtue of using $\tau$ is that its distribution completely decouples from the peaks constraint; the problematic correlation between $\{l_2,l_3\}$ and $\{y,z\}$ is absorbed into its definition (Appendix~\ref{app:mf}).
Figure~\ref{fig:Btau} further shows that, despite the scatter between $\tau$ and $q_{(5)}$ seen in Figure~\ref{fig:tauVsq5}, the protohalo densities do show a noticeable (but weaker) correlation with $\tau$.
In what follows, we therefore consider the barrier\footnote{We use conventions such that $\sig_{\rm 0T}$ is independent of redshift and the notation \delc\ in \eqn{tau-barrier} stands for $\delc(z)D(0)/D(z)$ (see discussion below equation~\ref{nuc-def}).}
\begin{align}
B &= \delc + \beta\,\sig_{\rm 0T}\,\tau\,\sqrt{(1-\gamma^2)/5}\notag\\
&\equiv \delc + \tilde\beta\,\sig_{\rm 0T}\,\tau\,,
\label{tau-barrier}
\end{align}
with $\beta$ being a constant and the \emph{only free parameter} in the model, and where the second line defines the (weakly) mass dependent quantity $\tilde\beta$. We will show results below for $\beta=0.5$, which we have found leads to an accurate description of the halo mass function at $z=0$ in our WMAP3 simulation. To test the robustness of the model, we then \emph{do not} change the value of $\beta$ when comparing with simulation results for the mass function and halo bias at different redshifts and/or for the Planck13 cosmology. We will refer to this model as `ESP$\tau$' below.

It is instructive to contrast our model for the barrier \eqref{tau-barrier} with that of \citet[][hereafter, PSD13]{psd13}, who considered the stochastic barrier
\be
B = \delc + \beta\,\sig_{\rm 0T}
\label{barrier-stochbeta}
\ee
whose slope $\beta$ \citep[motivated by the protohalo measurements of][]{Robertson2009} was taken to be a Lognormally distributed variable, independent of the shear and shape tensors. Both models lead to distributions of protohalo density at fixed mass that have approximately Lognormal shapes \citep[and therefore qualitatively agree with][]{Robertson2009}. In the PSD13 model, this is by construction. In the ESP$\tau$ model, on the other hand, the Lognormal-like shapes arise naturally since $\tau$ at fixed halo mass is close to being Chi-squared distributed with 5 degrees of freedom, which is close to the Lognormal shape \citep[see also][]{st02}.

Figure~\ref{fig:Bstoch} shows the stochasticity associated with the barrier \eqref{tau-barrier} in various forms. The left and middle panels show the distributions $p(\delta_{1\times}|m)$ of the density at first-crossing $\delta_{1\times}$ as predicted by excursion sets, which can be directly compared with measurements of protohalo density at fixed mass. In our model, this distribution is calculated by first evaluating $p(\tau|m) = p_5(\tau)n(m,\tau;\beta)/n(m;\beta)$, where $n(m,\tau;\beta)$ and $n(m;\beta)$ follow from the excursion set calculation (see below), and then changing variables using \eqn{tau-barrier}. The mean of $p(\delta_{1\times}|m)$ was shown as the solid curve in Figure~\ref{fig:Bq5}.

The left panel of Figure~\ref{fig:Bstoch} demonstrates that the Lognormal-like shape for this distribution is a robust prediction of the first-crossing (or up-crossing) constraint, and does not depend strongly on additionally imposing the peaks constraint. (In practice, the calculation without the peaks constraint is performed by setting $F(x)\to1$ in the results of section~\ref{sec:mf} and Appendix~\ref{app:mf}.) The middle panel compares the prediction of our peaks-based ESP$\tau$ model with protohalo measurements in the WMAP3 simulation. We see that the model consistently underpredicts the width of the distribution at all masses (a similar level of disagreement was also true of the PSD13 model). At the highest masses, however, it is likely that our $(200h^{-1}{\rm Mpc})^3$ WMAP3 simulation preferentially misses high density, large $q_{(5)}$ objects due to volume effects, leading to an increased mismatch in both the scatter as well as mean value of the distribution of $\delta_{1\times}$ (see also Figure~\ref{fig:Bq5}).

The right panel of Figure~\ref{fig:Bstoch} compares the prediction of $p(\tau|m)$ in the ESP$\tau$ model with the measured distributions of $\sqrt{5}q_{(5)}/\sqrt{1-\gam^2}$ for protohaloes in the WMAP3 simulation. We see that these distributions are very similar to each other (with a very weak mass dependence) and to the mass-independent distribution $p_5(\tau)$ which is related to the Chi-squared with 5 degrees of freedom (shown as the thick black dashed curve). This suggests that the disagreement seen in the middle panel is probably due to an over-simplified barrier shape. In principle, one could explore barrier models that are more complex nonlinear functions of $\tau$ \citep[see, e.g.,][]{cl01,st02}, or that involve other variables. We leave this to future work and restrict attention here to the simpler form in \eqn{tau-barrier}.

\begin{figure*}
\centering
\includegraphics[width=0.45\textwidth]{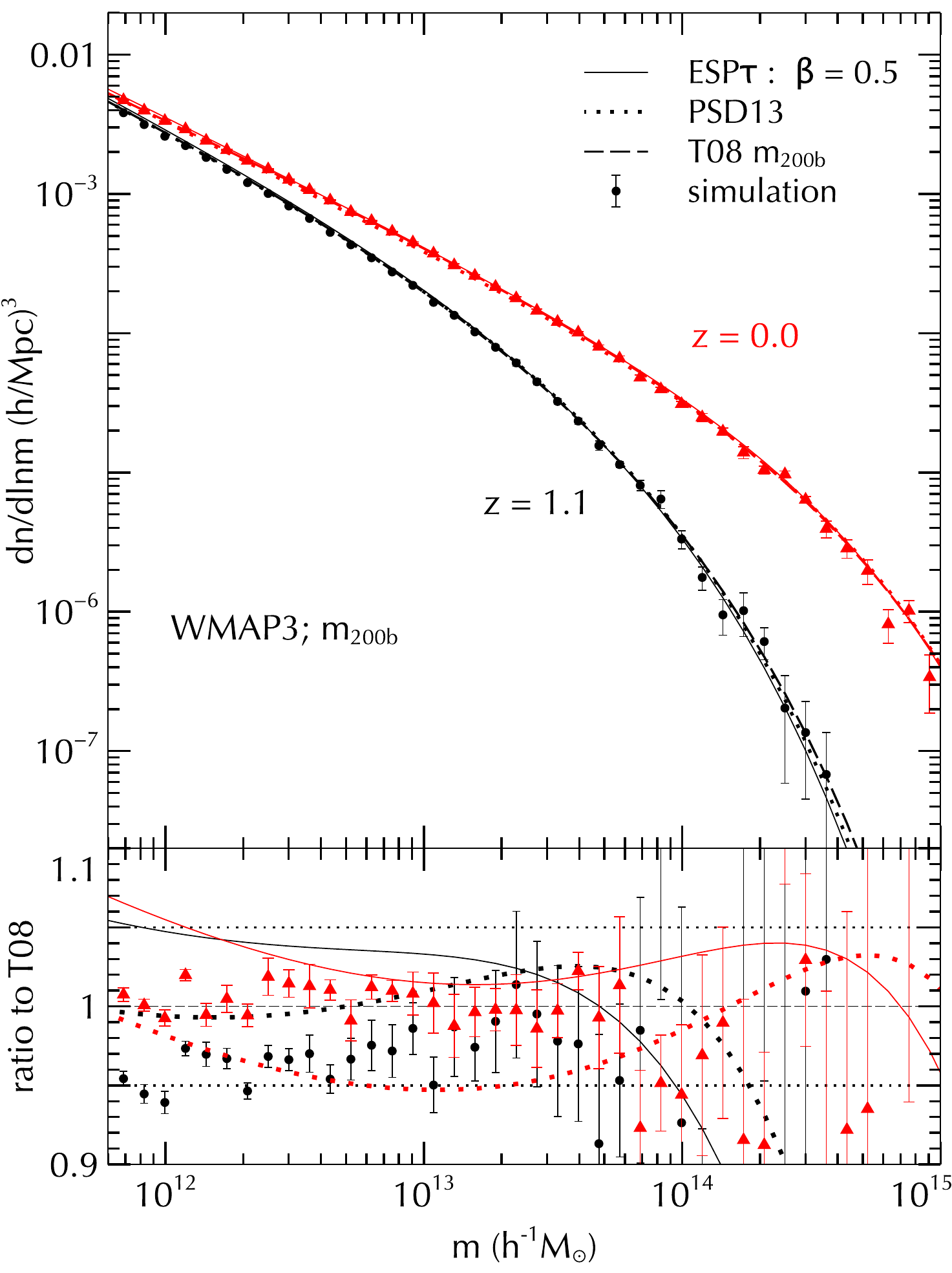}
\includegraphics[width=0.45\textwidth]{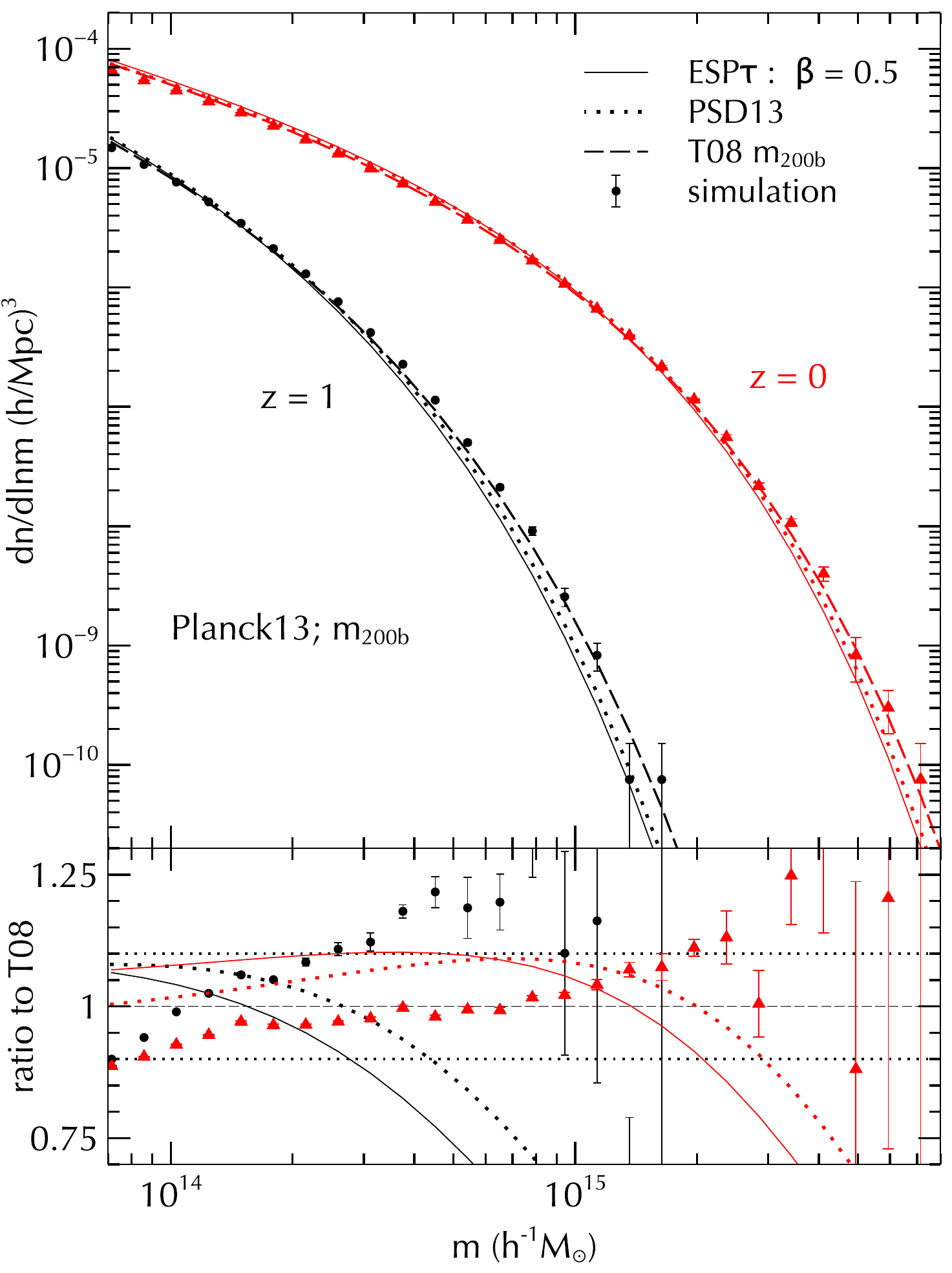}
\caption{Mass function given by \eqn{nmbeta} with $\beta=0.5$ (solid curves labelled $\textrm{ESP}\tau$) compared with measurements in $N$-body simulations (points with error bars) for the WMAP3 \emph{(left panel)} and Planck13 \emph{(right panel)} cosmologies, at two different redshifts. We also show the mass function from PSD13 (dotted curves) and the fitting function from \citep[][T08; dashed curves; we use the $\Delta=200$ parameter values from their Table 2, with a redshift dependence as described in their equations 5-8]{Tinker08}. Note that the Planck13 simulation used a $(2h^{-1}{\rm Gpc})^3$ box and probes the high-mass end of the mass function, while the WMAP3 simulation used a $(200h^{-1}{\rm Mpc})^3$ box and probes lower masses. All the simulation measurements used the mass definition $m_{\rm 200b}$ as described in the text. Each set of simulations had multiple realisations (see Table~\ref{tab:sims}), and the points show the mean counts over these runs for each bin while the error bars show the scatter around the mean. The lower panels show the ratio of the mass functions to the corresponding T08 curve. The horizontal dotted lines in the left (right) panel mark 5 (10) per cent deviations from the T08 fit.}
\label{fig:dnlowz}
\end{figure*}

\section{Halo mass function}
\label{sec:mf}
\noindent
In this section we present results for the halo mass function predicted by the ESP$\tau$ model, comparing them with measurements in $N$-body simulations and fits from the literature.

\subsection{Basic model}
\noindent
The calculation sketched in Appendix~\ref{app:mf} gives the mass function of excursion set peaks with the $\tau$-barrier \eqref{tau-barrier} as
\be
n(m;\beta) = \int_0^{\infty}\der\tau\, p_5(\tau)\,n(m,\tau;\beta)\,,
\label{nmbeta}
\ee
where we defined
\begin{align}
n(m,\tau;\beta) &\equiv \left|\frac{\der\ln\sig_{\rm 0T}}{\der\ln m}\right|\,\frac1{V_\ast}\,\frac{{\rm e}^{-(\nuc+\tilde\beta \tau)^2/2}}{\sqrt{2\pi}} 
\notag\\
&\ph{p()}
\times\int_0^\infty\frac{\der x\,F(x)}{\sqrt{2\pi(1-\gam^2)}}\,{\rm e}^{-\frac{\left(x-\gam\nuc-\gam\tilde\beta\tau\right)^2}{2(1-\gam^2)}}\notag\\
&\ph{p()\int\der x F(x)}
\times{\rm\bf ES}^\prime(x/\gam-\tilde\beta\tau)\,,
\label{nmtaubeta}
\end{align}
where $V_\ast=(6\pi)^{3/2}(\sig_{\rm 1G}/\sig_{\rm 2G})^3$, $F(x)$ is given by \eqn{Fbbks} and ${\rm\bf ES}^\prime$ is given by 
\be
{\rm\bf ES}^\prime(w) \equiv \frac{w}{2}\left[1+\erf{\frac{\Gam_\tau w}{\sqrt{2}}}\right] + \frac{{\rm e}^{-\Gam_\tau^2w^2/2}}{\sqrt{2\pi}\Gam_\tau}\,.
\label{ESprime}
\ee
with $\Gam_\tau = \gam/\tilde\beta$, and where $\tilde\beta$ was defined in \eqn{tau-barrier}. The integrals over $x$ and $\tau$ must be performed numerically; this is the same level of complexity that was present in the original excursion set peaks model of PSD13. 

\subsection{Low redshift results}
\noindent
Figure~\ref{fig:dnlowz} compares the ESP$\tau$ mass function of \eqn{nmbeta} with $\beta=0.5$ (solid curves) to measurements in our $N$-body simulations (points with error bars). The left panel shows measurements from the WMAP3 simulation which probes group-scale masses, while the right panel shows the larger volume Planck13 simulation which probes cluster-scale masses. We also show the mass function from the ESP model presented by PSD13 and the fit presented by \citet[][hereafter, T08]{Tinker08}, with the bottom panels showing the ratio with the T08 fit. 

As mentioned earlier, the value $\beta=0.5$ was chosen to ensure agreement between the ESP$\tau$ prediction and the WMAP3 simulation at $z=0$; we see that this is achieved at better than 5 per cent for $12 \lesssim \log_{10}\left(m/\Mh\right) \lesssim 14$. The same value of $\beta$ then continues to yield $\sim5$-$10$ per cent agreement with the WMAP3 simulation at $z\simeq1$ in a similar mass range, and at the $\sim 10$-$20$ per cent level in the Planck13 simulation at $z=0$ in the range $14.9 \lesssim \log_{10}\left(m/\Mh\right) \lesssim 15.5$. The level of agreement degrades at higher redshifts and masses, with the ESP$\tau$ model typically underpredicting the abundance of the rarest objects. These trends are clearly qualitatively very similar to those shown by the PSD13 model. Figure~\ref{fig:Bq5} suggests that this underprediction of rare objects could be related to the theoretical mean barrier height being substantially higher than the measured values of protohalo overdensity for the largest haloes.

\begin{figure}
\centering
\includegraphics[width=0.45\textwidth]{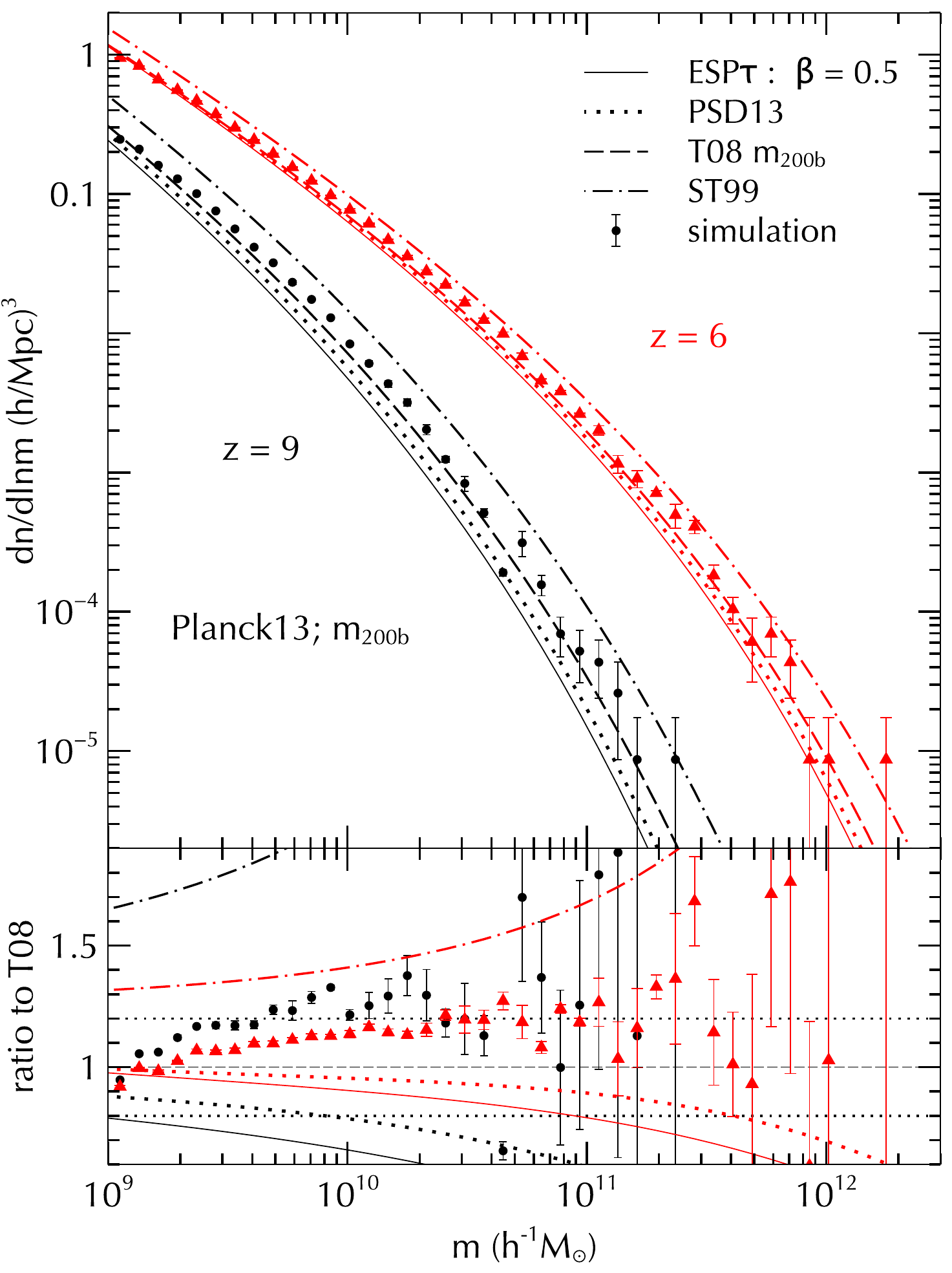}
\caption{Mass function comparison at high redshift. Formatting is the same as in Figure~\ref{fig:dnlowz}, except that we use the $(50h^{-1}{\rm Mpc})^3$ Planck13 simulation and additionally show the fitting function from \citet[][ST99]{st99}. The bottom panel again shows the ratio with the T08 fit, with the horizontal dotted lines indicating 20 per cent deviations. See text for a discussion.}
\label{fig:dnhiz}
\end{figure}

\subsection{High redshift results and non-universality}
\noindent
Figure~\ref{fig:dnhiz} shows a similar comparison using our high redshift Planck13 simulation. In addition to the models shown in Figure~\ref{fig:dnlowz}, we now also show the fit from \citet[][hereafter, ST99]{st99}. An important distinction between the ST99 fit and the other models is that this fit is \emph{universal}: the quantity $(m/\bar\rho)\der n/\der\ln \sig_{\rm 0T}$ in the ST99 fit depends only on the variable $\nu_{\rm c}$ (equation~\ref{nuc-def}). The PSD13 and ESP$\tau$ models, on the other hand, incorporate non-universal effects through the weak mass dependence of, primarily, the quantity $(V/V_\ast)$ and to some extent also that of $\gamma$, while the T08 fit (which was calibrated for the range $0\leq z \leq 2$) is non-universal due to the explicit redshift dependence of its parameters.

We see from the Figure that the T08 fit extrapolated to high redshifts continues to describe the $m_{\rm 200b}$ \textsc{Rockstar} halo mass function surprisingly well (at the $\sim10$-$20$ per cent level) in the range $9 \lesssim \log_{10}\left(m/\Mh\right) \lesssim 11$, which does not seem to have been emphasized in the literature previously. The ESP$\tau$ and PSD13 models now substantially underpredict the number of haloes, by $\sim20$ per cent at $z=6$ and by a factor of $\sim1.5$ or more at $z=9$, with ESP$\tau$ doing somewhat worse than PSD13. 

Other authors have previously explored the high redshift halo mass function \citep[see, e.g.,][]{Lukic+07,Watson+13,Despali+16} which is of great interest for studies of early structure formation, in particular the epoch of reionization. 
Figure~\ref{fig:dnhiz} highlights the fact that the high-redshift \textsc{Rockstar} mass function has a small but significant level of non-universality when using the $m_{\rm 200b}$ mass definition. To see this, note that the difference between the simulation data and the universal ST99 fit at fixed halo mass increases as we go from $z=6$ to $z=9$, and one can show that this increase \emph{cannot} be ascribed\footnote{For example, at $z=6$, the ST99 mass function is a factor $\sim1.3$ higher than the measurements in the range $10.5 \lesssim \log_{10}\left(m/\Mh\right) \lesssim 10.75$, corresponding to $2.77\lesssim\nuc\lesssim2.96$. At $z=9$, this range of $\nuc$ corresponds to the mass range $9 \lesssim \log_{10}\left(m/\Mh\right) \lesssim 9.3$, in which the ST99 mass function is a factor $\gtrsim1.5$ higher than the measurements, implying non-universality at the $\gtrsim15$ per cent level.} simply to a corresponding increase in the value of $\nu_{\rm c}$. The non-universal mass functions of ESP$\tau$, PSD13 and T08 also show identical trends, as expected.

\begin{figure*}
\centering
\includegraphics[width=0.9\textwidth]{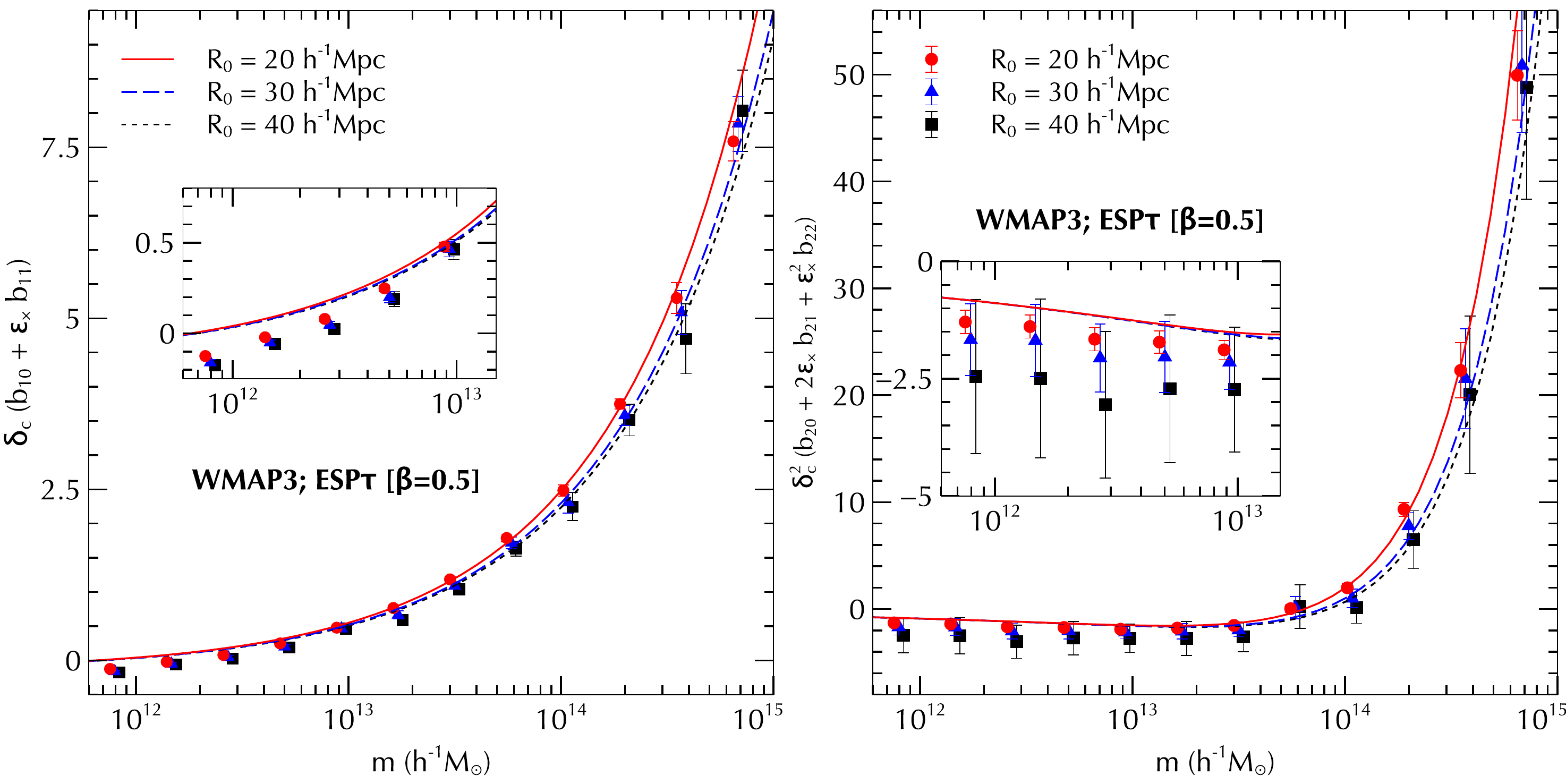}
\caption{Lagrangian linear \emph{(left panel)} and quadratic \emph{(right panel)} bias parameters $\delc b_1$ and $\delc^2b_2$, respectively, at $z=0$. The points with error bars show measurements surrounding $z=0$ CDM protohaloes in the initial conditions of the WMAP3 simulation at 3 different smoothing scales $R_0=20,30,40h^{-1}$Mpc (respectively, red circles, blue triangles and black squares). See text for details. The measurements are averaged over 9 realisations of the simulation, with error bars indicating the standard deviation around the mean. For clarity, we have given small horizontal offsets to points. Smooth curves show the scale-dependent prediction of the ESP$\tau$ model, with the solid red, dashed blue and dotted black curves corresponding to $R_0=20,30,40h^{-1}$Mpc, respectively. The insets in both panels zoom in on the low mass behaviour.}
\label{fig:b1-b2}
\end{figure*}

These results extend the conclusions drawn by T08 to higher redshifts than explored by those authors \citep[see also][]{Watson+13}. This discussion is particularly interesting in the context of recent results by \cite{Despali+16}, who showed that using the $m_{\rm vir}$ definition\footnote{Here $m_{\rm vir}$ is the mass enclosed in the virial radius $R_{\rm vir}$ at which the enclosed density reaches $\Delta_{\rm vir}\bar\rho$; the factor $\Delta_{\rm vir}(z)$ follows from solving the spherical collapse model in a $\Lambda$CDM background \citep{ecf96,bn98}, with $\Delta_{\rm vir}=18\pi^2$ for the Einstein-deSitter universe.} rather than $m_{\rm 200b}$ leads to a substantial \emph{decrease} in non-universality of spherical-overdensity haloes for $0\leq z\leq5$. Whether this continues to be the case at $z\lesssim9$ remains to be seen. In any case, Figure~\ref{fig:dnhiz} shows that the ESP$\tau$ and PSD13 models are \emph{too} non-universal, in that they under-predict halo abundances at fixed mass by increasingly larger factors at higher redshifts, even after accounting for the effect of increasing $\nu_{\rm c}$. We will comment on this in section~\ref{sec:conclude} in the context of constructing improved fits. 

\section{Halo bias}
\label{sec:bias}
\noindent
Traditionally, the concept of halo bias follows from the recognition that the small-scale, local density that drives gravitational collapse in some region of space is correlated with the density of the large scale environment of that region \citep{Kaiser84,bbks86,mw96,st99}. 

Recently, \cetal\ generalised this notion, arguing that halo bias is better thought of as the fact that the conditions that make the locations of collapse special (these could be related to more than just the small-scale density) are correlated with a number of variables characterising the large scale environment (these could be the density smoothed on different scales, or the large scale tidal field, and so on). For example, in addition to the density $\delta$, the slope of a random walk $\dot\delta=\der\delta/\der\sig_{\rm 0T}$ and peak curvature $x$ are important in determining halo abundances in models of excursion sets and/or peaks, and correlate strongly with large scale density, thereby leading to $k$-dependent bias in Fourier space \citep[][PSD13]{ps12a,ms12,mps12}. Correlating the locations of collapse with \emph{any} large-scale Gaussian variable that correlates with one or more of $\delta$, $\dot\delta$ and $x$ then leads to the \emph{same} $k$-dependent Fourier-space bias parameters.

\cetal\ further showed how this reinterpretation can be exploited in \emph{real} space to obtain \emph{model independent} estimates of the scale-independent coefficients of various scale-dependent terms in measurements of $n^{\rm th}$ order bias\footnote{This improves upon the model-dependent ``reconstruction'' technique \citep[][PSD13]{mps12} used previously by some of us \citep{psc+13}.}. Moreover, if the conditions determining collapse depend on variables that \emph{do not} correlate with the density in a random field (e.g., the rotational invariant $q_{(5)}^2$), then one expects new bias coefficients associated with these variables, possibly with their own scale dependence. In the following, we will apply some of the techniques suggested by \cetal\ to measure halo bias in our simulations (using the environments of protohaloes in the initial conditions) and compare the results with the Lagrangian predictions of the ESP$\tau$ and other models. 

\subsection{Scale dependent bias parameters $b_n$}
\noindent
Following \cite{mps12} and \cetal, a natural definition of the density bias parameters $b_n$ ($n=1,2,\ldots$) in real space is
\be
b_n \equiv \frac{\So^{n/2}}{\Sc^n}\avg{H_n(\del_0/\sqrt{\So})\,|\,\textrm{haloes}}\,,
\label{bn-def}
\ee
where $\So\equiv\avg{\delo^2}=\sig_{\rm 0T}^2(R_0)$ is the variance of the large scale density \delo,  $\Sc\equiv\avg{\delo\del}=\sig_{0\times,\textrm{TT}}^2(R,R_0)$ is the cross-correlation between the large scale and halo scale densities, and $H_n(x)\equiv{\rm e}^{x^2/2}(-\der/\der x)^n{\rm e}^{-x^2/2}$ are the ``probabilist's'' Hermite polynomials. For example, the linear bias parameter $b_1$ is then simply
\be
b_1 = \avg{\delo\,|\,\textrm{haloes}}/\avg{\delo\del}\,.
\label{b1-def}
\ee
Numerically the coefficients $b_n$ can be estimated by computing \delo\ on scale $R_0$ (centered at protohalo locations in an $N$-body simulation) and averaging the appropriate Hermite polynomial over specific bins in mass:
\be
\hat b_{n} = \frac{\So^{n/2}}{\Sc^n} \sum_{\alpha=1}^N H_n(\del_{0\alpha}/\sqrt{\So})/N\,,
\label{bn-estimator}
\ee
with $\del_{0\alpha}$ the measurement of \delo\ around the $\alpha$-th of $N$ haloes in the bin. 

\begin{figure*}
\centering
\includegraphics[width=0.9\textwidth]{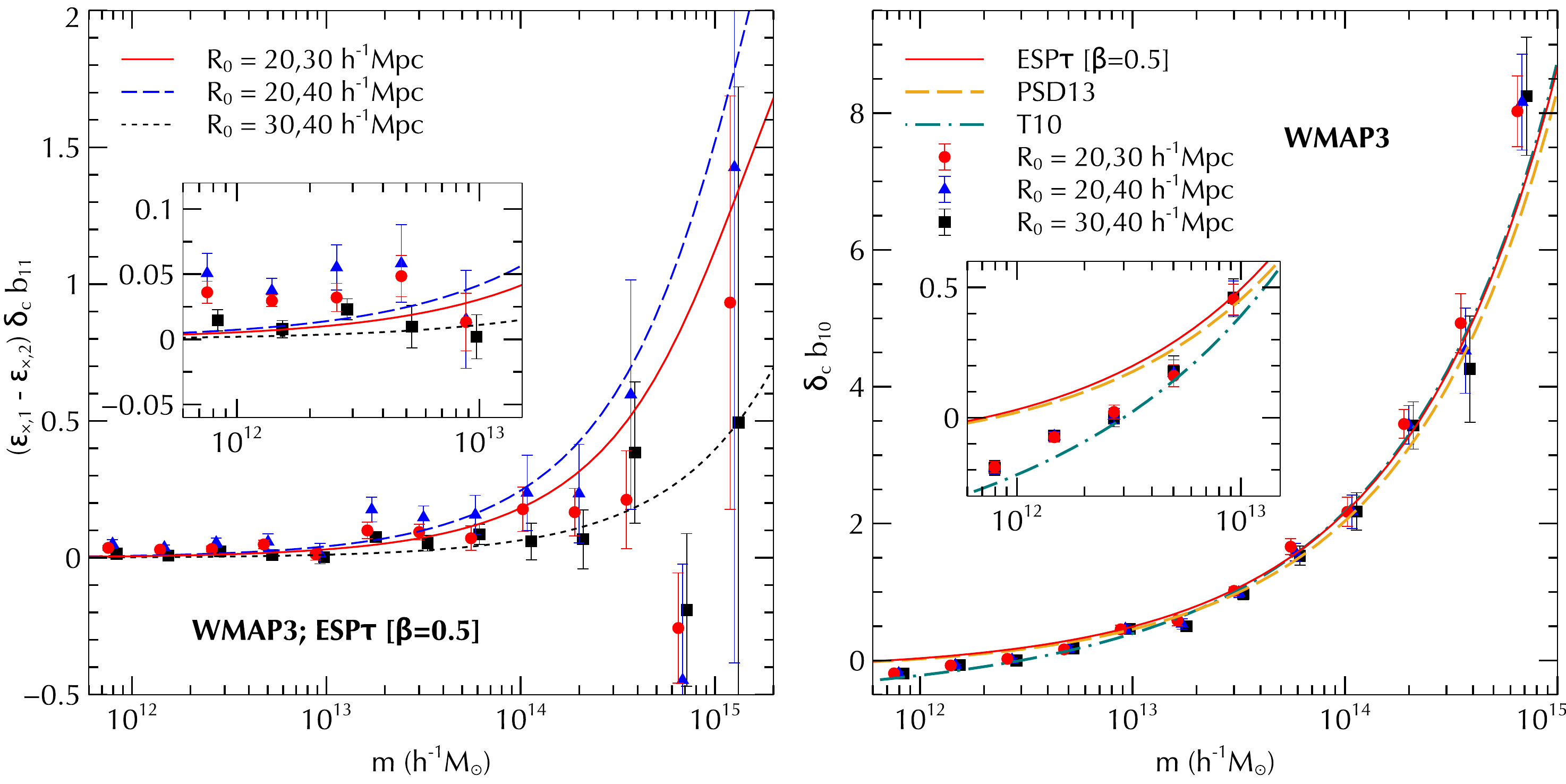}
\caption{Model-independent recovery of $b_{10}$ and $b_{11}$ from scale-dependent measurements of $b_1$. Both panels use pairwise combinations of the $b_1$ measurements at $R_0=20,30,40h^{-1}$Mpc shown in the left panel of Figure~\ref{fig:b1-b2}. As in that Figure, filled symbols show the mean of 9 realisations and error bars show the standard deviation around the mean. Red circles, blue triangles and black squares respectively show measurements using the pairs $(R_{0,(1)}$,$R_{0,(2)})=(20,30),(20,40),(30,40)h^{-1}$Mpc. \emph{(Left panel:)} Points show pairwise differences $\delc(b_{1(1)}-b_{1(2)})$ at scales $(R_{0,(1)}$,$R_{0,(2)})$. These are predicted to be equal to $\left(\epsilon_{\times,1}-\epsilon_{\times,2}\right)\delc b_{11}$, shown using the ESP$\tau$ model by the red solid, blue dashed and black dotted curves, respectively for the corresponding pairs of scales. \emph{(Right panel:)} Points show $\delc b_{10}$ recovered by inverting \eqn{linearbiasmatrix}. Smooth curves show the predictions of the ESP$\tau$ model (solid red), PSD13 model (orange dashed) and the fit to large-scale Fourier space measurements from \citet[][T10, dot-dashed green]{Tinker10}. The insets in each panel zoom in on the low mass behaviour.}
\label{fig:b11-b10}
\end{figure*}

Analytically, the calculation proceeds using the joint distribution of the excursion set peaks number density and \delo, and then computing the conditional expectation value $\avg{H_n(\delo/\sqrt{\So})|\textrm{ESP}\tau}$. We refer the reader to PSD13 for details \citep[see also][]{psc+13}. The result is
\begin{align}
b_n &= \sum_{r=0}^n\,\binom{n}{r}\,b_{nr}\epc^r
\label{bn-theory}
\end{align}
where the quantity \epc\ is defined as
\be
\epc\equiv s\,\sig_{1\times,\textrm{GT}}^2(R,R_0)/(\Sc\sig_{\rm 1m}^2)\,,
\label{epc-def}
\ee
with $s\equiv\avg{\del^2}=\sig_{\rm 0T}^2(R)$, and the scale-independent coefficients $b_{nr}$ are given by \eqn{bnr-def}. The Fourier-space scale dependence of bias mentioned above manifests through the presence of \epc\ in \eqn{bn-theory}. 

The points with error bars in Figure~\ref{fig:b1-b2} show measurements of $b_1$ (left panel) and $b_2$ (right panel) using the smoothed density field surrounding $z=0$ protohalo patches in the WMAP3 simulation, for three choices of smoothing scale. The smooth curves in each panel show the ESP$\tau$ predictions corresponding to each smoothing scale. We see that there is good agreement at high masses, whereas at low masses the model overpredicts both parameters \citep[although the low mass $b_2$ measurements are probably also being affected by finite volume effects, see, e.g.,][who saw similar effects using the ``reconstruction'' technique]{psc+13}.

\subsection{Scale independent bias coefficients $b_{nr}$}
\noindent
At large scales $R_0\gg R$, we have $\epc\to0$ and the bias coefficients reduce to $b_n\to b_{n0}$, which are the usual ``peak-background split'' bias coefficients routinely computed as derivatives of the halo mass function with respect to the spherical collapse threshold \delc\ \citep{mw96,mjw97,st99}. These are also directly related to traditional large scale measurements of bias in Fourier space \citep{ps12a,mps12}. The first two peak-background split coefficients in the ESP$\tau$ model are given in \eqn{b10b20-explicit}.

As \cetal\ pointed out, it is possible to separate out $b_{10}$ and $b_{11}$ from the combination $b_1=b_{10}+\epc b_{11}$ by measuring the latter using protohalo-centered averages of \emph{two} variables, such as the density at two different scales. If these measurements are denoted $\hat b_{1,(1)}$ and $\hat b_{1,(2)}$, then we can write the matrix equation
\be
\begin{pmatrix} \hat b_{1,(1)}\\ \hat b_{1,(2)} \end{pmatrix}
= 
\begin{pmatrix} 1 & \epsilon_{\times,1}\\ 1 & \epsilon_{\times,2} \end{pmatrix}
\begin{pmatrix} b_{10}\\ b_{11} \end{pmatrix}\,,
\label{linearbiasmatrix}
\ee
inverting which gives us estimates for both $b_{10}$ and $b_{11}$ with \emph{no assumptions} about the shape of the barrier or the variables affecting it. Similar arguments hold for higher order bias parameters, which can be separated with measurements using correspondingly larger numbers of variables.

Figure~\ref{fig:b11-b10} shows these reconstructions for linear bias using the $b_1$ measurements in the left panel of Figure~\ref{fig:b1-b2} pairwise. The points with error bars in the left panel show pairwise differences of these measurements, which are proportional to $b_{11}$, while those in the right panel show the values recovered for $b_{10}$ upon inverting \eqn{linearbiasmatrix}. While the $b_{11}$ measurements are rather noisy, we see that $b_{10}$ is recovered quite cleanly, with excellent agreement between the three sets of measurements. The smooth curves in each panel show various theoretical predictions and fits: in the left panel we show the predictions for $b_{11}$ of the ESP$\tau$ model, with the scale dependence arising from the proportionality factors which involve pairwise differences of \epc\ at different scales, while in the right panel we show the scale-independent predictions for $b_{10}$ of the ESP$\tau$ and PSD13 models, as well as the fitting function for $m_{\rm 200b}$ haloes from \citet[][T10]{Tinker10}. While the curves all agree with the high mass measurements, at low masses the measurements clearly prefer the T10 fit.

\subsection{Moments of the stochastic barrier from bias measurements}
\label{subsec:meanbarrier}
\noindent
We have seen above that the ideas presented in \cetal\ lead to model-independent measurements of linear and higher order Lagrangian bias coefficients. These measurements can also, in principle, be extended to obtain model-independent estimates of the moments of the barrier distribution itself, as we discuss next.

The key ingredient here is the existence of certain ``consistency relations'' between the scale-independent bias coefficients $b_{nr}$ \citep[][PSD13]{mps12}. A simple derivation outlined in Appendix~\ref{app:bias} shows that the $b_{nr}$ must obey \eqn{sumbnr}; turning this around leads to expressions for the moments of the barrier. For example, at linear order we find the mean relation
\be
\avg{B/\sig_{\rm 0T}\,|m} = \sig_{\rm 0T} \left(b_{10}+b_{11}\right)\,,
\label{avgBm}
\ee
which is just a consequence of the fact that the cross correlation we used to measure bias parameters is equivalent to the enclosed mean density profile in Lagrangian space. At quadratic order, the consistency relations give the variance
\begin{align}
\textrm{Var}\left(B/\sig_{\rm 0T}\,|m\right) &= 1 + \sig_{\rm 0T}^2\bigg\{b_{20}+2b_{21}+b_{22}\notag\\
&\ph{1 + \sig_{\rm 0T}^2\left[b_{20}\right]}
-\left(b_{10}+b_{11}\right)^2\bigg\}\,,
\label{varBm}
\end{align}
and so on for higher order moments. Measurements of the bias coefficients can therefore lead to an alternative description of the distribution of barrier heights. This was exploited by \cetal, who used Monte Carlo random walks to show that the first two consistency relations hold, for both a simple spherical collapse model and a stochastic model which resembles the one in \eqn{tau-barrier}. In Appendix \ref{app:wtdwalks}, we generalize these results of \cetal\ to the case of peak weighted walks with stochastic barriers. 

\begin{figure}
\centering
\includegraphics[width=0.45\textwidth]{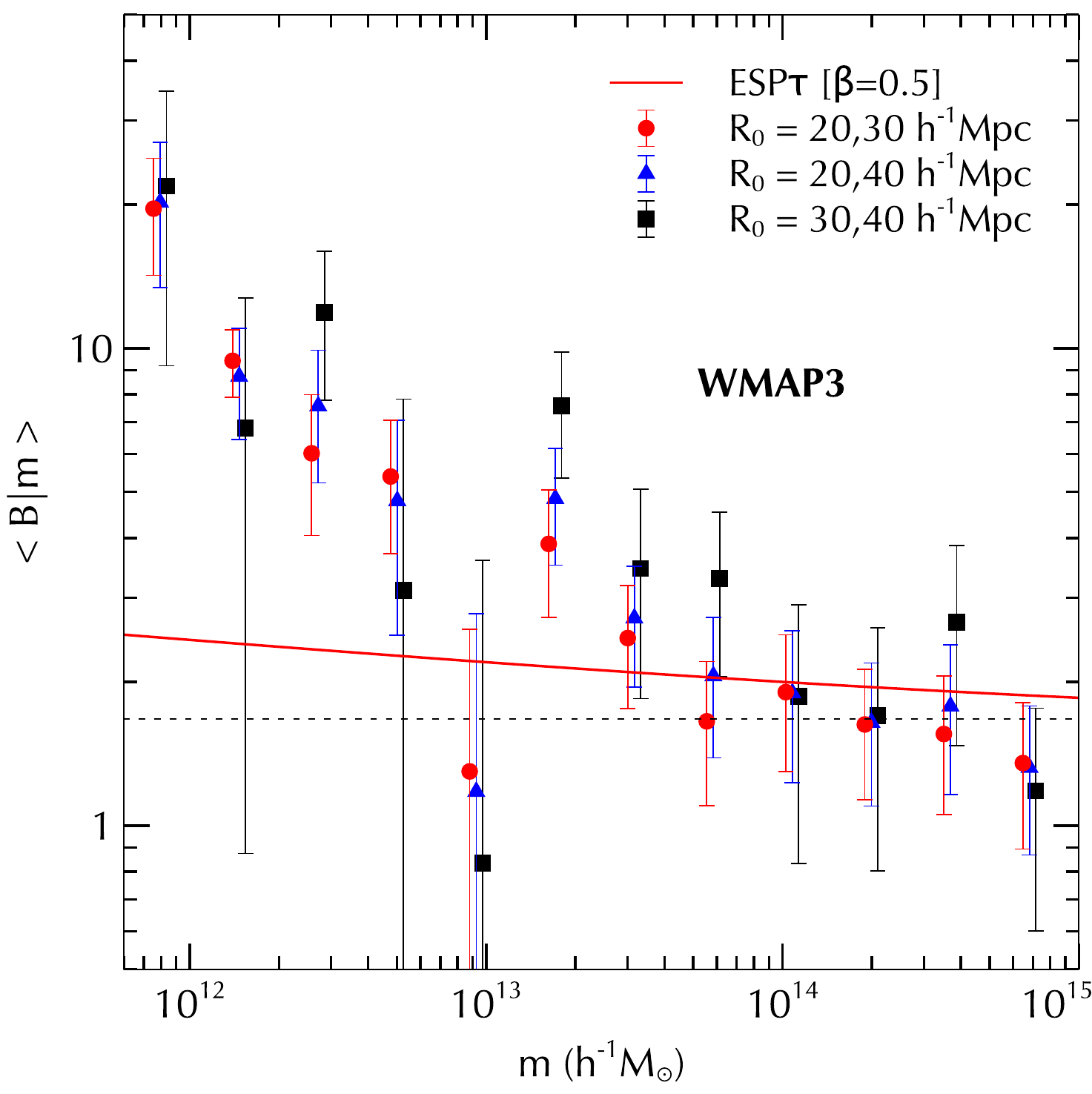}
\caption{Model-independent measurement of the mean barrier $\avg{B|m}$ recovered by applying \eqn{avgBm} to the measurements of $b_{10}$ and $b_{11}$ in Figure~\ref{fig:b11-b10}.}
\label{fig:meanbarrier}
\end{figure}

In an $N$-body simulation, however, these measurements are rather susceptible to noise and finite volume effects. Figure~\ref{fig:meanbarrier} shows the mean barrier $\avg{B|m}$ recovered from the $b_{10}$ and $b_{11}$ measurements shown in Figure~\ref{fig:b11-b10}. The smooth solid curve shows the theoretical expression from the ESP$\tau$ model, which was also shown in Figure~\ref{fig:Bq5}. Comparing Figures~\ref{fig:Bq5} and~~\ref{fig:meanbarrier} we see that, apart from the noise at all masses, the new measurements substantially overestimate the mean protohalo densities of small mass objects (note that the vertical axis in Figure~\ref{fig:meanbarrier} has a logarithmic scale). We can trace this effect to the overestimation of the pairwise differences of bias parameters shown in the left panel of Figure~\ref{fig:b11-b10}, which is quite likely due to finite volume effects similar to those affecting the $b_2$ measurements of Figure~\ref{fig:b1-b2}. These overestimated differences are then divided by factors involving pairwise differences of \epc, whose values are much smaller than unity at these masses, leading to the overestimate of the mean barrier in Figure~\ref{fig:meanbarrier}. It will be interesting to repeat this analysis on simulations with larger volume at the same mass resolution, an exercise we leave for future work.

\subsection{Nonlocal bias induced by shear}
\noindent
\label{subsec:nonlocalbias}
As a final application of ideas presented by \cetal, we explore the bias parameters $c_n$ associated with the fact that shear is important in determining the locations of collapse. Whereas the parameters $b_n$ could be recovered through protohalo-centered averages over variables that correlate with density, the $c_n$ analogously require averages over variables that correlate with shear. And since the distribution of $q_{(5)}^2$ (or $\tau^2$) at random locations is Chi-squared rather than Gaussian, the Hermite polynomials are replaced with generalized Laguerre polynomials, which are the orthogonal polynomials associated with the Chi-squared distribution. 

The presence of nonlocal shear-induced bias in the Eulerian (i.e., gravitationally evolved) spatial distribution of haloes is of considerable interest due to its effect on the halo bispectrum and, consequently, its potential effect on the recovery of cosmological parameters and the detection of primordial non-Gaussianity from large scale structure \citep{chan/scoccimarro/sheth:2012,baldauf/seljak/etal:2012,scs13,Saito+14,bcdp14,afsv15}.
Previous authors \citep[][hereafter, SCS13]{chan/scoccimarro/sheth:2012,baldauf/seljak/etal:2012,scs13} have argued that nonlocal terms in the \emph{Lagrangian} field would also be relevant for Eulerian measurements of bias, since gravitational evolution induces a coupling between density and shear. The prediction of these Lagrangian terms in the analytical ESP$\tau$ model is therefore potentially of great interest for interpreting Eulerian measurements of the halo bispectrum, and represents a qualitative departure from the model of PSD13 in which the barrier stochasticity was an \emph{ad hoc} ingredient that did not correlate with the protohalo envionment.

The numerical estimator for nonlocal bias suggested by \cetal\ is based on the large-scale rotational invariant $Q_{(5)}$ defined at scale $R_0$ exactly like $q_{(5)}$ is defined at the halo scale $R$, and is given by
\be
\hat c_{2j} = (-1)^j\,r_\zeta^{-2j} \frac1N\,\sum_{\alpha=1}^N L_j^{(3/2)}(5Q_{(5)\alpha}^2/2)
\label{c2j-estimator}
\ee
with $j=1,2,\ldots$, where $Q_{(5)\alpha}$ is the measurement of $Q_{(5)}$ around the $\alpha$-th of $N$ protohaloes in the bin, $L_j^{(\alpha)}(x)\equiv x^{-\alpha}{\rm e}^x(\der/\der x)^j({\rm e}^{-x}x^{j+\alpha})/j!$ are the generalized Laguerre polynomials and $r_\zeta$ is the correlation coefficient between the large-scale and small-scale shear (see Appendix~\ref{app:nonlocalbias} for details).

\begin{figure}
\centering
\includegraphics[width=0.45\textwidth]{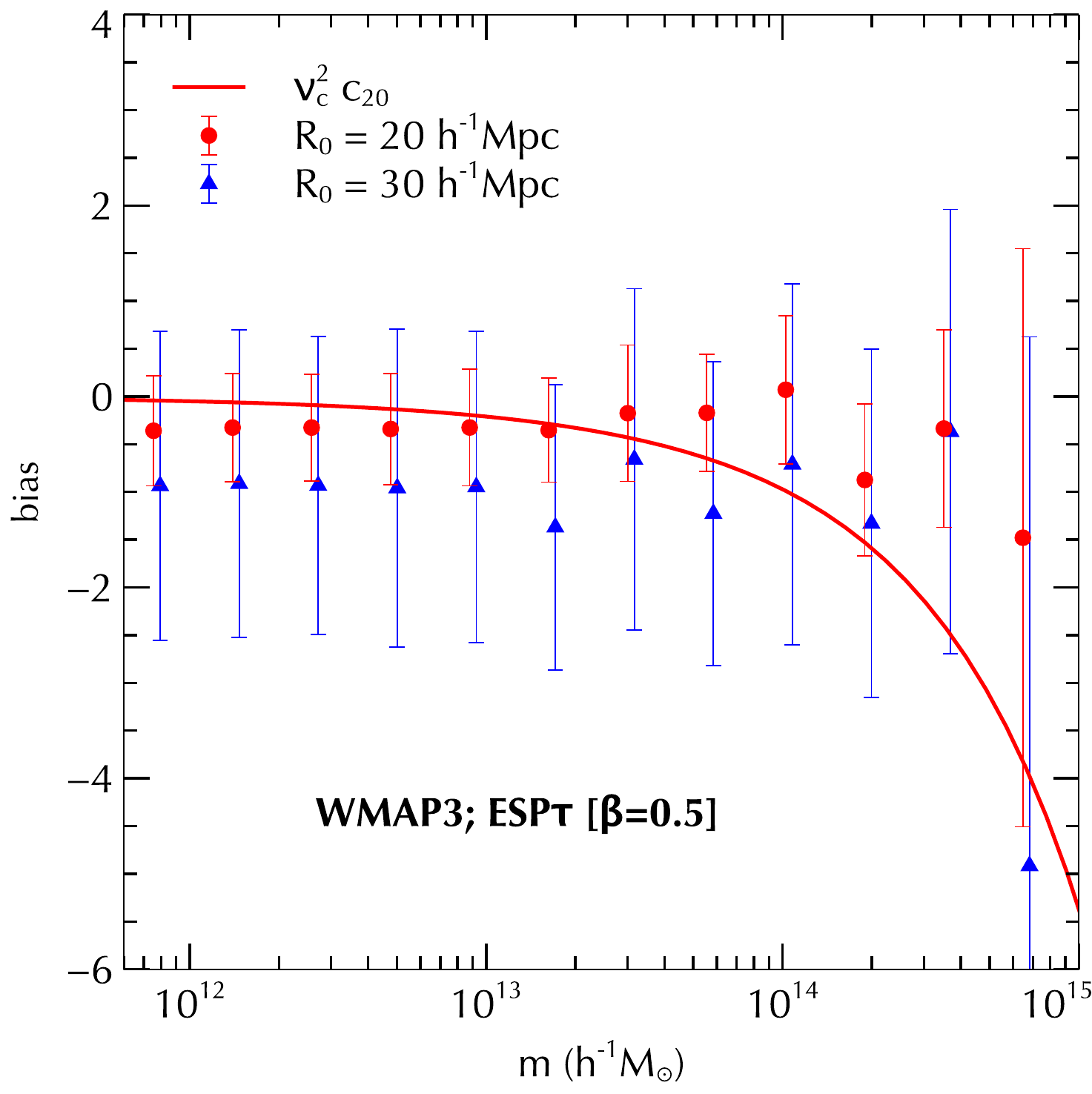}
\caption{Lagrangian shear bias $c_2$. Points with error bars show measurements using large scale shear in the initial conditions surrounding $z=0$ protohaloes in the WMAP3 simulation at scales $R_0=20h^{-1}$Mpc (red circles) and $30h^{-1}$Mpc (blue triangles). See text for details. Points show the mean over 9 realisations and error bars show the standard deviation around the mean. Solid red curve shows the scale-independent prediction $c_{20}$ in the ESP$\tau$ model. For ease of comparison with SCS13, we plot $\delc^2c_2^{\rm (SCS13)}\equiv\nuc^2c_2$.}
\label{fig:c2}
\end{figure}

As \cetal\ showed, the scale-dependent coefficients $c_{2j}$ are expected to have a structure very similar to that of $b_{n}$ (equation~\ref{bn-theory}). For example, at lowest order $j=1$, one has $c_2 = c_{20} + 2\tilde\epsilon_\times c_{21} + \tilde\epsilon_\times^2c_{22}$, where $\tilde\epsilon_\times$ is a coefficient analogous to \epc\ (equation~\ref{epc-def}), now involving shear rather than density. At very large scales, the only relevant term is $c_{20}$ (similarly, $c_{\{2j\}0}$ at order $j$). In Appendix~\ref{app:nonlocalbias}, we show that, in our analytical model which uses $\tau$ rather than $q_{(5)}$, this is very well approximated by
\begin{align}
c_{\{2j\}0}(m;\beta)
&\simeq \frac{1}{j!}\times\frac{1}{n(m;\beta)}\,\left(\frac{\beta^2}{2}\right)^j\, \left(\frac{1}{\beta}\frac{\p}{\p\beta}\right)^j\,n(m;\beta)\,,
\label{c2j-theory-alt}
\end{align}
which is precisely analogous to the definition of the peak-background split density bias coefficients as derivatives of the mass function with respect to \delc. As \cetal\ showed using Monte Carlo experiments, and as  further discussed in Appendix \ref{app:wtdwalks}, the scale dependent pieces $c_{21}$ and $c_{22}$ are not so important at the scales we probe in our simulations; given the size of the errorbars, these can be safely neglected. 

Figure~\ref{fig:c2} compares our protohalo measurements using \eqn{c2j-estimator} at two large scales with the expectation from \eqn{c2j-theory}. To connect with the results of SCS13, we note that their convention for the nonlocal bias coefficient they called $c_2$ is related to ours through $\delc^2c_2^{\rm (SCS13)}=\nuc^2c_2$, which is the combination we display. Unfortunately, the measurements are dominated by noise, with hints of the same scale-dependent finite volume effects seen in the $b_2$ measurements of Figure~\ref{fig:b1-b2}. While there is broad agreement between the measurements and the model prediction, drawing sharper conclusions would require better error control (see also the previous discussion of the mean barrier). Nevertheless, Figure~\ref{fig:c2} represents the first attempt at directly measuring Lagrangian nonlocal bias in $N$-body simulations.

\section{Summary \& Discussion}
\label{sec:conclude}
\noindent
The role of tidal influences or shear in the formation of dark matter haloes has long been recognised as an important ingredient in models of hierarchical structure formation \citep{bm96,smt01,st02,blp14}. In parallel, analytical work on peaks theory and the excursion set approach with correlated steps has shown how these two can be combined \citep{ms12,ps12b}, leading to realistic and accurate ``excursion set peak'' (ESP) models of the halo mass function and scale dependent halo bias \citep[][PSD13]{psd13}. This ESP approach, however, does not incorporate the tidal influences mentioned above, a situation that we have rectified in this paper.

We have introduced an ESP model that explicitly depends on the initial tidal field by including the effect of the latter in the critical density threshold for gravitational collapse. Although simplified, this collapse barrier (equation~\ref{tau-barrier}) is well-motivated from standard ellipsoidal collapse arguments which typically lead to an enhancement of the critical density by terms depending on the rotational invariants of the tidal tensor  \citep{smt01,st02,scs13}. Moreover, by adjusting the value of a single parameter (the strength $\beta$ of the tidal effect in the collapse barrier), the resulting `ESP$\tau$' model gives an excellent description of the halo mass function (Figure~\ref{fig:dnlowz}) and linear and quadratic scale-dependent bias parameters (Figure~\ref{fig:b1-b2}) at low redshift over a wide range of halo masses. The performance of the ESP$\tau$ model in this regard is comparable to that of the PSD13 model. 

Our model highlights the intimate connection between scale dependent halo bias and the moments of the stochastic halo formation barrier (section~\ref{subsec:meanbarrier}; Figure~\ref{fig:meanbarrier}), recently discussed by \citet[][\cetal]{cps17}\footnote{Along the way, we also employed the techniques proposed by \cetal\ to estimate the coefficients $b_{10}$ and $b_{11}$ of the linear bias parameter in a \emph{model independent} way, comparing these with predictions of the ESP models and fits from the literature (Figure~\ref{fig:b11-b10}). Our analytical results were validated using Monte Carlo experiments involving peak weighted random walks, a novel technique that we describe in Appendix~\ref{app:wtdwalks}.}.
Moreover, the presence of shear in the ESP$\tau$ collapse barrier \eqref{tau-barrier} leads to qualitatively new features that the PSD13 model lacks. First, the Lognormal-like shape of the distribution of overdensities of protohalo patches is a \emph{prediction} of this model, rather than being an assumption as in the case of PSD13. 
Quantitatively, this prediction provides a reasonably good description of direct measurements of protohalo overdensities in $N$-body simulations (Figures~\ref{fig:Bq5} and~\ref{fig:Bstoch}, although see below). 

Secondly, the fact that shear is important in determining the locations of collapse leads to conceptually new bias coefficients (\cetal) that are \emph{not predicted} in the PSD13 model. We presented an estimate (the first in the literature) of this `nonlocal' Lagrangian bias in Figure~\ref{fig:c2}; although noisy, the measurements are in broad agreement with the ESP$\tau$ prediction. Another manifestation of this nonlocal bias arises due to the gravitational coupling between density and shear, an effect which has been discussed in the literature and recently measured in simulations (see section~\ref{subsec:nonlocalbias} for references). Our Lagrangian prediction will have consequences for the interpretation of this Eulerian effect, which in turn can affect cosmological parameter recovery and the hunt for primordial non-Gaussianity from large scale structure.

By extrapolating the models to high redshifts $6\leq z\leq 9$ and comparing with simulations, we also demonstrated  that the ESP models predict a level of non-universality in the mass function that is substantially larger than what is measured (Figure~\ref{fig:dnhiz})\footnote{A simple extrapolation of the non-universal fitting function by  \citet{Tinker08} to these redshifts, on the other hand, leads to a surprisingly accurate description of the simulation results.}. This is one of several features in our model that deserves more attention: for example, the measurements of the scale independent linear bias coefficient $b_{10}$ are significantly overpredicted by the model at low masses (right panel of Figure~\ref{fig:b11-b10}), the mean value of the protohalo overdensity distribution is also overpredicted (Figure~\ref{fig:Bq5}), while its scatter is substantially underpredicted (Figure~\ref{fig:Bstoch}).

There are at least two possible approaches to resolving these issues. A phenomenological solution might involve direct comparisons of the mass function and barrier stochasticity with measurements in simulations at several redshifts and masses, with the purpose of fitting the parameter $\beta$ and/or introducing more parametrized terms (e.g., nonlinear terms in the shear) in the barrier \eqn{tau-barrier}. This approach would have the immediate benefit of providing well-motivated, self-consistent and accurate fitting forms for the mass function and nonlinear, nonlocal bias.

A second, potentially more rewarding approach, could involve exploring other variables that can be important in determining the locations of collapse. For example, recent work on the formation histories of CDM haloes by  \citet{blp14} has demonstrated two very interesting facts: first, the ellipsoidal collapse model of \citet{bm96}, after accounting for the nontrivial initial shapes of protohalo patches, tends to systematically overpredict collapse times \citep[a similar conclusion was also reached indirectly by][through an analysis of warm dark matter simulations]{hp14}, and secondly, the overdensities of protohaloes of a given mass strongly correlate with their actual collapse times in the simulation. 

Since collapse times, as well as their intimate connection with other properties such as halo concentration, are natural statistical ingredients in the excursion set formalism \citep{lc93,cs13}, developing an accurate physical model of halo formation times and their dependence on protohalo shape and initial shear is clearly of great interest. In addition to potentially resolving (at least some of) the issues faced by the ESP$\tau$ model of this paper, such a model would also have consequences for accurate predictions of the concentration-mass-redshift relation \citep{Ludlow+14,Ludlow+16} and measured trends of halo assembly bias \citep{st04,gsw05}. 

One point of interest in this regard is that the peaks framework naturally predicts an assembly bias signature from the correlation between the peak curvature and the large scale density \citep{dwbs08}.  \citet{ms14-markov} made the point that the effect is more general: in the excursion set framework the assembly bias effect is due to the correlation between the slope of the density profile and the large scale density - the peaks constraint is not necessary.  In our excursion set peaks approach, coefficients like $b_{11}$ essentially measure the strength of both correlations. In addition, \citet{cs13} showed that stochasticity in halo formation is another source of assembly bias, and \cetal\ show how stochasticity and slope correlations combine in coefficients like $b_{11}$.  So it follows that these coefficients contain information about assembly bias (at least for the unevolved, Lagrangian field). In principle, then, measuring these coefficients could lead to a new window for detecting assembly bias.
Exploring these ideas is the subject of work in progress.

\section*{Acknowledgements}
EC, AP and RKS thank E. Sefusatti and L. Guzzo for their kind hospitality at INAF-OAB Merate during the summer of 2014. AP and OH gratefully acknowledge use of computing facilities at ETH, Z\"urich, where part of this work was completed. The research of AP is supported by the Associateship Scheme of ICTP, Trieste and the Ramanujan Fellowship awarded by the Department of Science and Technology, Government of India.

\bibliography{masterRef}

\appendix

\section{Details of calculations}
\label{app:details}

\subsection{Correlations}
\label{app:corrs}
\noindent
The covariances between the matrix elements are given by
\begin{align}
\avg{\psi_{ij}\psi_{kl}} &= \frac{\sig_{\rm 0T}^2}{15}\left(\del_{ij}\del_{kl}+\del_{ik}\del_{jl}+\del_{il}\del_{jk}\right)\,,
\label{psi-psi}\\
\avg{\zeta_{ij}\zeta_{kl}} &= \frac{\sig_{\rm 2G}^2}{15}\left(\del_{ij}\del_{kl}+\del_{ik}\del_{jl}+\del_{il}\del_{jk}\right)\,,
\label{zeta-zeta}\\
\avg{-\psi_{ij}\zeta_{kl}} &= \frac{\sig_{\rm 1m}^2}{15}\left(\del_{ij}\del_{kl}+\del_{ik}\del_{jl}+\del_{il}\del_{jk}\right)\,,
\label{psi-zeta}\\
\avg{\eta_i\eta_j} &= \frac{\sig_{\rm 1G}^2}{3}\del_{ij}\,,
\label{eta-eta}\\
\avg{\eta_i\psi_{jk}} &= 0 = \avg{\eta_i\zeta_{jk}}\,,
\label{eta-cross}
\end{align}
where $\del_{ij}$ is the Kronecker delta, and $\sig_{\rm 1m}^2$ was defined below \eqn{gam-def}.

The variables defined in equations~\eqref{xyz-def} and~\eqref{nul2l3-def} have the following non-zero covariances:
\begin{align}
\avg{x^2} = 1 = \avg{\nu^2} &\quad;\quad \avg{x\,\nu} = \gam\notag\\
15\avg{y^2} = 1 = \avg{l_2^2} &\quad;\quad \avg{y\,l_2} = \gam/\sqrt{15}\notag\\
5\avg{z^2} = 1 = \avg{l_3^2} &\quad;\quad \avg{z\,l_3} = \gam/\sqrt{5}
\label{diag-cov}
\end{align}
where $\gam$ was defined in \eqn{gam-def}. Additionally, 
\begin{align}
&\sig_{\rm 0T}^{-2}\avg{\psi_A^2} = 1/15 = \sig_{\rm 2G}^{-2}\avg{\zeta_A^2}\notag\\ 
&(\sig_{\rm 0T}\sig_{\rm 2G})^{-1}\avg{\psi_A\,\zeta_B} = -(\gam/15)\delta_{AB}\,,
\label{offdiag-cov}
\end{align}
where $A,B=4,5,6$.

\subsection{Mass function for ESP with shear}
\label{app:mf}
\noindent
The differential number density of peaks can be written as \citep{vdwb96}
\begin{align}
&\Cal{N}_{\rm pk}(\nu,l_2,l_3,q_{(3)},x,y,z)\,\der\nu\,\der l_2\,\der l_3\,\der q_{(3)}\,\der x\,\der y\,\der z\notag\\
&\ph{p}
=\frac1{V_\ast} \times \sqrt{\frac2\pi}\,\frac{3q_{(3)}^2(\sqrt{3}\der q_{(3)})}{(1-\gam^2)^{3/2}}{\rm e}^{-3q_{(3)}^2/2(1-\gam^2)}\notag\\ 
&\ph{p}
\times \frac{\der\nu\,{\rm e}^{-\nu^2/2}}{\sqrt{2\pi}}
\times\frac{\der x\,{\rm e}^{-(x-\gam\nu)^2/2(1-\gam^2)}}{\sqrt{2\pi(1-\gam^2)}}\notag\\
&\ph{p}
\times\frac{3^2\cdot5^{5/2}}{\sqrt{2\pi}}\der y\,\der z\,F(x,y,z)\,{\rm e}^{-15y^2/2}\,{\rm e}^{-5z^2/2}\chi_{\zeta}(x,y,z)\notag\\
&\ph{p}
\times \frac{\der l_2\,\der l_3}{2\pi(1-\gam^2)}{\rm e}^{-[(l_2-\gam\sqrt{15}y)^2+(l_3-\gam\sqrt{5}z)^2]/2(1-\gam^2)}\,,
\label{Npk-generic}
\end{align}
where $V_\ast=(6\pi)^{3/2}(\sig_{\rm 1G}/\sig_{\rm 2G})^3$, the constraints on the shape eigenvalues are given by
\be
\chi_{\zeta}(x,y,z)=\HT(y)\HT(y-z)\HT(y+z)\HT(x-3y+z)\,,
\label{chi_zeta}
\ee 
and, following \citet[][hereafter, BBKS]{bbks86}, we denote
\be
F(x,y,z)\equiv(x-2z)y(y^2-z^2)[(x+z)^2-(3y)^2]\,.
\label{Fxyz}
\ee
Introducing the variables 
\begin{align}
q &\equiv\sqrt{3}q_{(3)}/\sqrt{1-\gam^2}\notag\\
{\tilde l}_2 &\equiv (l_2-\gam\sqrt{15}\,y)/\sqrt{1-\gam^2}\notag\\
{\tilde l}_3 &\equiv (l_3-\gam\sqrt{5}\,z)/\sqrt{1-\gam^2}\,, 
\label{ql2till3til-def}
\end{align}
\eqn{Npk-generic} can be rewritten as
\begin{align}
&\Cal{N}_{\rm pk}(x,y,z,\nu,\tilde l_2,\tilde l_3,q)\,\der x\,\der y\,\der z\,\der\nu\,\der \tilde l_2\,\der \tilde l_3\,\der q\notag\\
&=\frac1{V_\ast} \times \der q\,p_3(q)\times \frac{\der \tilde{l}_2\,\der \tilde{l}_3}{2\pi}{\rm e}^{-(\tilde{l}_2^2+\tilde{l}_3^2)/2}\notag\\
&\times \frac{\der\nu\,{\rm e}^{-\nu^2/2}}{\sqrt{2\pi}}\times\frac{\der x\,{\rm e}^{-(x-\gam\nu)^2/2(1-\gam^2)}}{\sqrt{2\pi(1-\gam^2)}}\notag\\
&\times\frac{3^2\cdot5^{5/2}}{\sqrt{2\pi}}\frac{\der y\,\der z}{1-\gam^2}\,F(x,y,z)\,{\rm e}^{-15y^2/2}\,{\rm e}^{-5z^2/2}\chi_{\zeta}(x,y,z)\,,
\label{Npk-generic-2}
\end{align}
where $p_3(q)=\sqrt{2/\pi}\,q^2\,{\rm e}^{-q^2/2}$. The first line of \eqref{Npk-generic-2} (apart from the factor $1/V_\ast$) simply gives the conditional distribution $\der q\,\der\tilde l_2\,\der\tilde l_3\,p(q,\tilde l_2,\tilde l_3|\zeta_A=0)$ of the variables $\{q,\tilde{l}_2,\tilde{l}_3\}$ in the eigenbasis of $\zeta_{ij}$.
As we see, this distribution \emph{completely decouples from the peaks constraint}. We can now apply a 3-dimensional polar coordinate transformation $\{q,\tilde l_2,\tilde l_3\}\to\{\tau,\theta_\tau,\phi_\tau\}$, the radial coordinate being $\tau$ as defined in \eqn{tau-def} which also reads $\tau^2=q^2+\tilde{l}_2^2+\tilde{l}_3^2$. Assuming that the \emph{excursion set constraint} only depends on $\tau$, the angular part of this transformation can be integrated over: due to the form of $p_3(q)$ this involves the integrals $\int_0^\pi\der\theta_\tau\,\sin^3\theta_\tau=4/3$ and $\int_0^\pi\der\phi_\tau\,\cos^2\phi_\tau=\pi/2$, where the range of integration of $\phi_\tau$ follows from noting the restriction $q\geq0$. We are left with
\begin{align}
&\Cal{N}_{\rm pk}(x,y,z,\nu,\tau)\,\der x\,\der y\,\der z\,\der\nu\,\der \tau\notag\\
&=\frac1{V_\ast} \times \der \tau\,p_5(\tau) \times \frac{\der\nu\,{\rm e}^{-\nu^2/2}}{\sqrt{2\pi}}
\times\frac{\der x\,{\rm e}^{-(x-\gam\nu)^2/2(1-\gam^2)}}{\sqrt{2\pi(1-\gam^2)}}\notag\\
&\times\frac{3^2\cdot5^{5/2}}{\sqrt{2\pi}}\frac{\der y\,\der z}{1-\gam^2}\,F(x,y,z)\,{\rm e}^{-15y^2/2}\,{\rm e}^{-5z^2/2}\chi_{\zeta}(x,y,z)\,,
\label{Npk-tau}
\end{align}
where $p_5(\tau)$ was defined in \eqn{p5tau-def} and does not couple to the peaks constraint.

The mass function of excursion set peaks (``haloes'') is given by multiplying the peaks number density by the excursion set constraint and integrating over all relevant variables:
\be
n(m) = \left|\frac{\der\ln\sig_{\rm 0T}}{\der\ln m}\right|\,\int\Cal{D}{\Xv}\,\Cal{N}_{\rm pk}(\Xv)\,{\rm\bf ES}(\sig_{\rm 0T},\{\del,B\})\,,
\label{dnESP-formal}
\ee
where the excursion set constraint, restricted to up-crossing of a barrier $B$ by the walk height \del, is given by
\be
{\rm\bf ES}(\sig_{\rm 0T},\{\del,B\}) = (\dot\del-\dot B)\,\HT(\dot\del-\dot B)\,\dir(\nu-B/\sig_{\rm 0T})\,,
\label{ES-constraint}
\ee
with the overdot denoting a derivative with respect to $\sig_{\rm 0T}$, and where\Xv\ in our case represents $\{\nu,\tau,x,y,z\}$. 

We use a barrier of the form \eqref{tau-barrier} for the reasons discussed in the text. To proceed, we need the derivative of the barrier \eqref{tau-barrier}. Ignoring the weak mass dependence of \gam, this is given by
\be
\dot B = \tilde\beta\,\tau + v\quad;\quad v\equiv \tilde\beta\,\sig_{\rm 0T}\,\dot \tau\,,
\label{Bdot}
\ee
where $\tilde\beta$ was defined in \eqn{tau-barrier}, and where we introduced a new stochastic variable $v$ whose distribution can be shown to be Gaussian with mean zero and variance $\tilde\beta^2/\gam^2$, \emph{independent} of all the other variables at the halo scale \citep[][although see below]{ms14-stoch}. The quantity $\dot\del$ on the other hand is strongly correlated with the peak curvature $x$. In fact, for Gaussian filtering we have $\dot\del = x/\gam$ (a special case of $\gam\dot\psi_{ij}=-\zeta_{ij}/\sig_2$). Assuming this relation and integrating over the Gaussian distribution of $v$ leads to the replacement 
\be
{\rm\bf ES}(\sig_{\rm 0T},\{\del,B\}) \to 
{\rm\bf ES}^\prime(x/\gam-\tilde\beta \tau)\,\dir(\nu-\tilde\beta \tau-\nuc)\,,
\label{ES-replaced}
\ee
where \nuc\ was defined in \eqn{nuc-def} and ${\rm\bf ES}^\prime$ in \eqn{ESprime}. We then use \eqns{Npk-tau} and~\eqref{ES-replaced} in \eqn{dnESP-formal} and integrate over $\{y,z\}$ exactly like in BBKS. Since these variables don't enter the excursion set constraint, this is equivalent to using the variables $\Xv=\{\nu,\tau,x\}$ in \eqn{dnESP-formal}, with
\begin{align}
&\Cal{N}_{\rm pk}(\nu,\tau,x)\,\der \nu\,\der \tau\,\der x\notag\\
&=\frac{\der \tau\,p_5(\tau)}{V_\ast}\, \frac{\der\nu\,{\rm e}^{-\nu^2/2}}{\sqrt{2\pi}}
\,\frac{\der x\,F(x)\,\HT(x)}{\sqrt{2\pi(1-\gam^2)}}\,{\rm e}^{-(x-\gam\nu)^2/2(1-\gam^2)}
\label{Npk-shear-tau}
\end{align}
where
\begin{align}
F(x)&=\frac12\left(x^3-3x\right)\left\{\erf{x\sqrt{\frac52}}+\erf{x\sqrt{\frac58}}
  \right\} \notag\\
&\ph{x^3-3x}
+ \sqrt{\frac2{5\pi}}\bigg[\left(\frac{31x^2}{4}+\frac85\right){\rm
    e}^{-5x^2/8} \notag\\
&\ph{\sqrt{x^3-3x+\frac2{5\pi}}[]}
+ \left(\frac{x^2}{2}-\frac85\right){\rm
    e}^{-5x^2/2}\bigg]\,,
\label{Fbbks}
\end{align}
(equation~A15 of BBKS). Finally, integrating over the Dirac-delta in $\nu$ leads to \eqn{nmbeta}.

\subsection{Scale-independent bias coefficients}
\label{app:bias}
\noindent
The scale-independent bias coefficients $b_{nr}$ ($n=1,2,\ldots$; $0\leq r\leq n$) in the ESP$\tau$ model are given by
\begin{align}
\delc^n b_{nr} &= (-1)^r\,\sum_{j=0}^{n-r}\,\binom{n-r}{j}\,\avg{\mu_j\lambda_{n-j}|\textrm{ESP}\tau}\,,
\label{bnr-def}
\end{align}
where the functions $\mu_j$ and $\lam_j$ are given by
\begin{align}
\mu_j(\nuc,\tau;\beta) &\equiv \nuc^j\,H_j(\nuc+\tilde\beta \tau)
\label{muj-def}\\
\lambda_j(\nuc,\tau,x;\beta) &\equiv (-\Gam\nuc)^j\,H_j\left(\frac{x - \gam\nuc - \gam\tilde\beta\tau}{\sqrt{1-\gam^2}}\right)\,,
\label{lamj-def}
\end{align}
with $\Gam\equiv\gam/\sqrt{1-\gam^2}$, and the conditional average $\avg{g|\textrm{ESP}\tau}$ of a function $g(\tau,x)$ at excursion set peaks in this model is
\be
\avg{g|\textrm{ESP}\tau}=\frac{\int\der \tau\,\der x\,\Cal{N}_{{\rm ESP}\tau}(\nuc,\tau,x;\beta)\,g(\tau,x)}{\int\der \tau\,\der x\,\Cal{N}_{{\rm ESP}\tau}(\nuc,\tau,x;\beta)}\,,
\label{<g(x,tau)|ESP>}
\ee
where
\be
\Cal{N}_{{\rm ESP}\tau}(\nuc,\tau,x;\beta)=\Cal{N}_{\rm pk}(\nuc+\tilde\beta \tau,\tau,x)\,{\rm\bf ES}^\prime(x/\gam-\tilde\beta\tau)
\label{NESPxtau}
\ee
with $\Cal{N}_{\rm pk}(\nu,\tau,x)$ given by \eqn{Npk-shear-tau}, while for a function $g(\tau)$ this becomes
\be
\avg{g|\textrm{ESP}\tau} = \frac{\int\der \tau\,p_5(\tau)\,n(m,\tau;\beta)\,g(\tau)}{\int\der \tau\,p_5(\tau)\,n(m,\tau;\beta)}\,,
\label{<g(tau)|ESP>}
\ee
where $n(m,\tau;\beta)$ was defined in \eqn{nmtaubeta}. The first two peak-background split coefficients evaluate to
\begin{align}
\delc b_{10} &= \avg{\mu_1|\textrm{ESP}\tau} + \avg{\lam_1|\textrm{ESP}\tau}\notag\\
\delc^2 b_{20} &= \avg{\mu_2|\textrm{ESP}\tau} + 2\avg{\mu_1\lam_1|\textrm{ESP}\tau} + \avg{\lam_2|\textrm{ESP}\tau}\,.
\label{b10b20-explicit}
\end{align}
Of particular interest from the point of view of barrier stochasticity are the ``consistency relations'' first noticed by \cite{mps12} and discussed by PSD13 and \cetal: summing over \eqn{bnr-def}, we can easily see that\footnote{To prove the second equality in \eqn{sumbnr}, one can exchange the order of summation and recognize the binomial expansion of $(1-1)^{n-j}=\delta_{nj}$.}
\begin{align}
&\sum_{r=0}^n\,\binom{n}{r}\,\delc^nb_{nr} \notag\\
&= \sum_{r=0}^n\sum_{j=0}^{n-r}\,\binom{n}{r}\binom{n-r}{j}\,(-1)^r\,\avg{\mu_j\lam_{n-j}|\textrm{ESP}\tau}\notag\\
&=\avg{\mu_n|\textrm{ESP}\tau}\notag\\
&=\nuc^n\avg{H_n\left(B/\sig_{\rm 0T}\right)|\textrm{ESP}\tau}\,.
\label{sumbnr}
\end{align}
We discuss this relation further in the main text.

\subsection{Nonlocal bias induced by shear}
\label{app:nonlocalbias}
\noindent
To obtain the analytical analogue of the bias \eqref{c2j-estimator} associated with shear, we must define the large scale version of $\tau$, which we denote $T$. If the components of the large-scale shear tensor are organised into the variables $\{\delo/\sqrt{\So},L_2,L_3,L_4,L_5,L_6\}$ exactly like the small-scale shear was organized into $\{\nu,l_2,l_3,l_4,l_5,l_6\}$ (except that we standardize using $\sig_{\rm 0T}(R_0)=\sqrt{\So}$ instead of $\sig_{\rm 0T}(R)=\sqrt{s}$), then we can define
\begin{align}
3Q_{(3)}^2 &\equiv L_4^2+L_5^2+L_6^2\notag\\
Q &\equiv\sqrt{3}Q_{(3)}/\sqrt{1-\gam_\times^2}\notag\\
{\tilde L}_2 &\equiv (L_2-\gam_\times\sqrt{15}\,y)/\sqrt{1-\gam_\times^2}\notag\\
{\tilde L}_3 &\equiv (L_3-\gam_\times\sqrt{5}\,z)/\sqrt{1-\gam_\times^2}\,, 
\label{QL2tilL3til-def}
\end{align}
which are analogous to $\{q,\tilde l_2,\tilde l_3\}$ defined in \eqn{ql2till3til-def}, and since we are still working in the eigenbasis of the \emph{small-scale} shape tensor $\zeta_{ij}$, the relevant correlation coefficient that appears is $\gam_\times$ defined by
\be
\gam_\times \equiv \frac{\textrm{Cov}(L_A,\zeta_A)}{\sqrt{\textrm{Var}(L_A)\textrm{Var}(\zeta_A)}} = \frac{\sig_{1\times,\textrm{GT}}(R,R_0)^2}{\sig_{\rm 0T}(R_0)\sig_{\rm 2G}(R)}\,.
\label{gamx-def}
\ee
Analogously to $\tau$ (equation~\ref{tau-def}), we can then define $T$ using
\be
T^2\equiv Q^2 + \tilde L_2^2 + \tilde L_3^2\,.
\label{T-def}
\ee
Turning to the cross-correlation between shear at different scales, we note that in the absence of any constraint on the small-scale reference frame (and keeping in mind that the $l_A$ and $L_A$ are all standardized), the relevant correlation coefficient between $\tau$ and $T$ would have been $r$ defined by
\be
r \equiv \textrm{Cov}(l_A,L_A) =  \frac{\sig_{0\times,\textrm{TT}}^2(R,R_0)}{\sig_{\rm 0T}(R)\sig_{\rm 0T}(R_0)} = \frac{S_\times}{\sqrt{sS_0}}\,.
\label{r-def}
\ee
Due to the requirement of being in the $\zeta_{ij}$ eigenbasis, this changes to $r_\zeta$ given by
\begin{align}
r_\zeta & = (r - \gam\gam_\times)/\sqrt{(1-\gam^2)(1-\gam_\times^2)}\,.
\label{r_zeta}
\end{align}
The joint distribution of $\tau$ and $T$ is then $\der\tau\,\der T\,p(\tau, T)=\der\tau\,p_5(\tau)\,\der T\,p_5(T|\tau;r_\zeta)$ where $p_5(\tau)$ is given by \eqn{p5tau-def} and we have (using the expression for a bivariate Chi-squared distribution with $5$ degrees of freedom)
\begin{align}
&p_5(T|\tau;r_\zeta)\,\der T \notag\\
&= \der T\,T\,\frac{{\rm e}^{-(T^2+r_\zeta^2\tau^2)/2(1-r_\zeta^2)}}{(1-r_\zeta^2)}\left(\frac{T}{r_\zeta \tau}\right)^{3/2} I_{3/2}\left(\frac{r_\zeta \tau T}{1-r_\zeta^2}\right)\,,
\label{pT|tau}
\end{align}
with $I_{3/2}(x)$ a modified Bessel function of the first kind. 

Strictly speaking, what we require for the calculation of shear-induced bias is the conditional distribution $p(T|\tau,\dot\tau)$, since the excursion set constraint depends on both $\tau$ and its derivative. Although $\dot\tau$ does not correlate with any of the variables at the halo scale, it \emph{does} correlate with large scale shear. 
At leading order, keeping track of the additional correlation with $\dot\tau$ leads to a scale dependent bias $c_2$ involving coefficients $c_{20}$, $c_{21}$ and $c_{22}$, exactly analogous to the structure of $b_2$ (equation~\ref{bn-theory}). The coefficient $c_{20}$ is given in full by
\begin{align}
c_{20} 
&= -\int\frac{\der\tau\,\der x\,\der v\,\der\nu}{n(m;\beta)}\,\Cal{N}_{\rm pk}(\nu,\tau,x)\,p(v)\,{\rm\bf ES}(\sig_{\rm 0T},\{\del,B\})
\notag\\
&\ph{-\int\times}
\times\left[L_1^{(3/2)}(\tau^2/2)+  \Gamma^2 \tau v /\tilde{\beta} + \Gamma^2 L^{(-1)}_1(\Gamma^2 v^2/\tilde{\beta}^2) \right] \notag\\
& = -\avg{L_1^{(3/2)}(\tau^2/2)+ \Gamma^2 \tau v /\tilde{\beta} + \Gamma^2 L^{(-1)}_1(\Gamma^2 v^2/\tilde{\beta}^2)|\textrm{ESP}\tau} \,,
\label{c2j-theory}
\end{align}
%
%
which can be compared with the expression for $b_{20}$ in \eqn{b10b20-explicit}. The full derivation of $c_{20}$ in \eqn{c2j-theory} can be found in \cetal\ (their Appendices A and B). Since the $\textrm{ESP}\tau$ constraint and the barrier in \eqn{tau-barrier} do not change the correlation structure of $p(T|\tau,\dot\tau)$, $c_{20}$ can be computed from equation B5 of \cetal, replacing $q$ with $\tau$, $\dot{q}$ with $v$, and adding an extra $F(x)$ weighting.

As \cetal\ showed, however, the pieces in \eqn{c2j-theory} involving the variable $v$ -- which arise from the correlation between $\dot\tau$ and the large scale $T$ -- are much smaller than the first piece involving only the Laguerre polynomial in $\tau$ (see also Figure \ref{fig:c2fromwalks}). This first piece can be derived by approximating $p(T|\tau,\dot\tau)\simeq p(T|\tau)=p_5(T|\tau;r_\zeta)$, as we show below. At order $j$, we have
\begin{align}
&(-1)^j\,r_\zeta^{2j}\,c_{2j,0}\notag\\
&\simeq\avg{L_j^{(3/2)}(T^2/2)|\textrm{ESP}\tau, \,\avg{T \dot\tau}  \simeq 0)}\notag\\ 
&= \int\der T\,L_j^{(3/2)}(T^2/2)\,p(T|\textrm{ESP}\tau, \,\avg{T \dot\tau}  \simeq 0)\notag\\
&= \int\der T\,L_j^{(3/2)}(T^2/2)\,\int\der \tau\,p(T|\tau)\,p(\tau|\textrm{ESP}\tau)\notag\\
&= \int\der T\,L_j^{(3/2)}(T^2/2)\,\int\der \tau\,p_5(T|\tau;r_\zeta)\,\frac{p_5(\tau)n(m,\tau;\beta)}{n(m;\beta)}\notag\\
&= \int\der \tau\,\frac{p_5(\tau)n(m,\tau;\beta)}{n(m;\beta)}\int\der T\, p_5(T|\tau;r_\zeta)\,L_j^{(3/2)}(T^2/2)\notag\\
&= \int\der \tau\,\frac{p_5(\tau)n(m,\tau;\beta)}{n(m;\beta)}\,r_\zeta^{2j}\,L_j^{(3/2)}(\tau^2/2)\notag\\
&= r_\zeta^{2j}\,\avg{L_j^{(3/2)}(\tau^2/2)|\textrm{ESP}\tau}\,,
\label{avg-LjT|ESP}
\end{align}
where the last equality but one uses a special case of $\int\der T\,p_{2(\alpha+1)}(T|\tau;r_\zeta)L_j^{(\alpha)}(T^2/2) = r_\zeta^{2j}L_j^{(\alpha)}(\tau^2/2)$ for a conditional Chi-squared distribution with $2(\alpha+1)$ degrees of freedom (obtained by setting $3/2\to\alpha$ in the last two terms in equation~\ref{pT|tau}), which can be proved using the identity 7.421(4) of \cite{grad-ryzh}. 
Using the definition of the generalized Laguerre polynomials and integrating the last line but one in \eqn{c2j-theory}  by parts, we see that this term can be written as in \eqn{c2j-theory-alt}.

\section{Weighted random walks}
\label{app:wtdwalks}
\noindent
The procedure to generate quasi-random walks with a given correlation structure was described for the first time in a cosmological context by \citet{bcek91}. This technique was then used by \cite{ms12,mps12} and later by many others to test the accuracy, within the up-crossing approximation, of analytic predictions for the first crossing rate and the bias parameters with respect to the outcome of numerical excursion set experiments.  

In this Appendix we describe the key steps to generate Monte Carlo walks with \emph{peak weights}, which is a straightforward extension of the algorithm of \citet{bcek91} to include the effects of performing the excursion set calculation at peaks of the initial density field.
\begin{figure}
\centering
\includegraphics[width=0.45\textwidth]{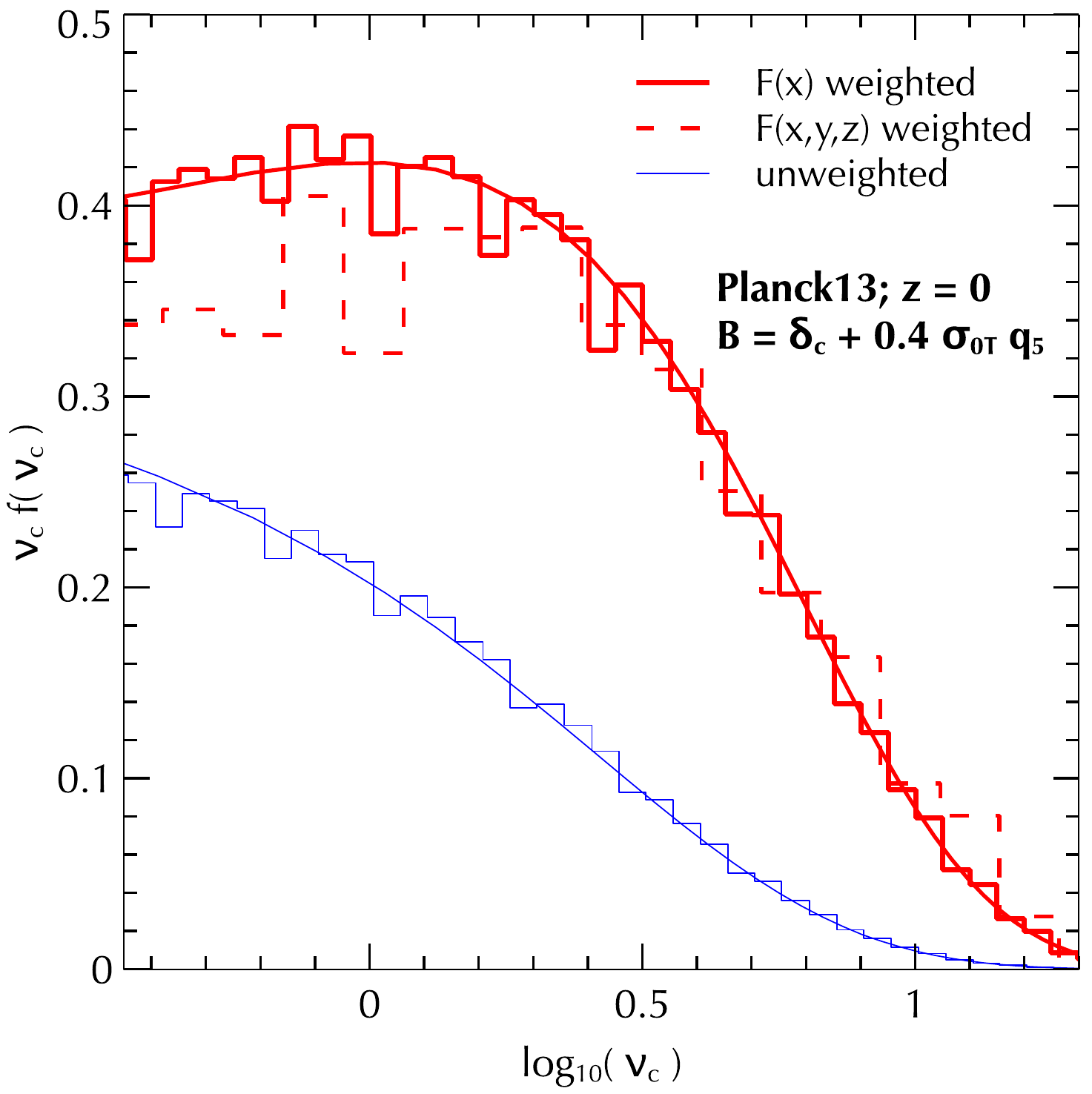}
\caption{Results using peak-weighted random walks. First crossing rate of the barrier $\delta_c + 0.4\, \sigma_{\rm 0T}\, q_5$ (histograms). Different linestyles correspond to different weighting schemes: thick solid red are weighted using \eqn{eq:pkwt}, dashed red by replacing $F(x)\HT(x)\to F(x,y,z)\chi_{\zeta}(x,y,z)$ in that equation, and thin solid blue are unweighted. Smooth curves show the corresponding analytical predictions using the ESP$\tau$ model. (Here, the unweighted case simply amounts to setting $(V/V_\ast)F(x)\to1$ in equation~\ref{nmtaubeta}).}
\label{fig:fnufromwalks}
\end{figure}
Given a set of linear Fourier modes of the density field, $\delta_{k_i}$, for each random walk and at each step labelled by its smoothing scale $R$, the density field convolved with a window function $W(k R)$ can be computed as
\begin{equation}
\delta_R = \sum_i \delta_{k_i} W(k_i R)\,.
\end{equation}
For simplicity, and without loss of generality, we assumed the walk is centered at the origin of coordinates. 

A certain number of variables, e.g., the smoothing scale and the value of density field,  are stored at the scale where the first crossing condition, say $\delta_R \ge B$, is satisfied, and the same is done at the larger scales one later wants to use to compute the bias parameters, using the estimators in \eqn{bn-estimator}. The barrier $B$ of \eqn{tau-barrier} is constructed by using an appropriate number of independent Gaussian random numbers to build Chi-squared objects like $q_5$ and $\tau$. For the relevant steps in the random walk, we also store the value of $x$, defined in \eqn{xyz-def}, 
\begin{equation}
x_R = \sum_i k_i^2 \delta_{k_i} W_G(k_i R) /\sigma_{2G}(R)\,.
\end{equation}
The peak weight is computed as 
\be
w_{\rm pk}(x_R,R) = \left(V/V_\ast\right)\, F(x_R)\, \HT(x_R)\,,
\label{eq:pkwt}
\ee
where $V=(4\pi/3)R^3$, $V_\ast$ was defined below \eqn{nmtaubeta}, $F(x)$ is given by \eqn{Fbbks}, and $\HT(x)$ is the Heaviside theta function that enforces $x>0$. When computing the first crossing distribution, this weight is used as the contribution of each walk to the frequency histogram. If $N_i$ walks first cross the barrier in the $i^{\rm th}$ bin of smoothing scale, then the corresponding weighted first crossing fraction of walks $f_i$ is
\be
f_i = \sum_{\alpha=1}^{N_i} w_{\rm pk}(x_{R_\alpha},R_\alpha)/N_{\rm tot}\,,
\ee
with $\alpha$ counting over the walks contributing to this bin, and where $N_{\rm tot}$ is the total number of walks simulated. The first crossing distributions of walks with and without weighting are shown in Figure~\ref{fig:fnufromwalks}, along with analytical up-crossing based results. We see that the analytical results provide an excellent description of the numerical measurements in each case, thus validating the main technical approximation underlying the ESP formalism.

\begin{figure*}
\centering
\includegraphics[width=0.8\textwidth]{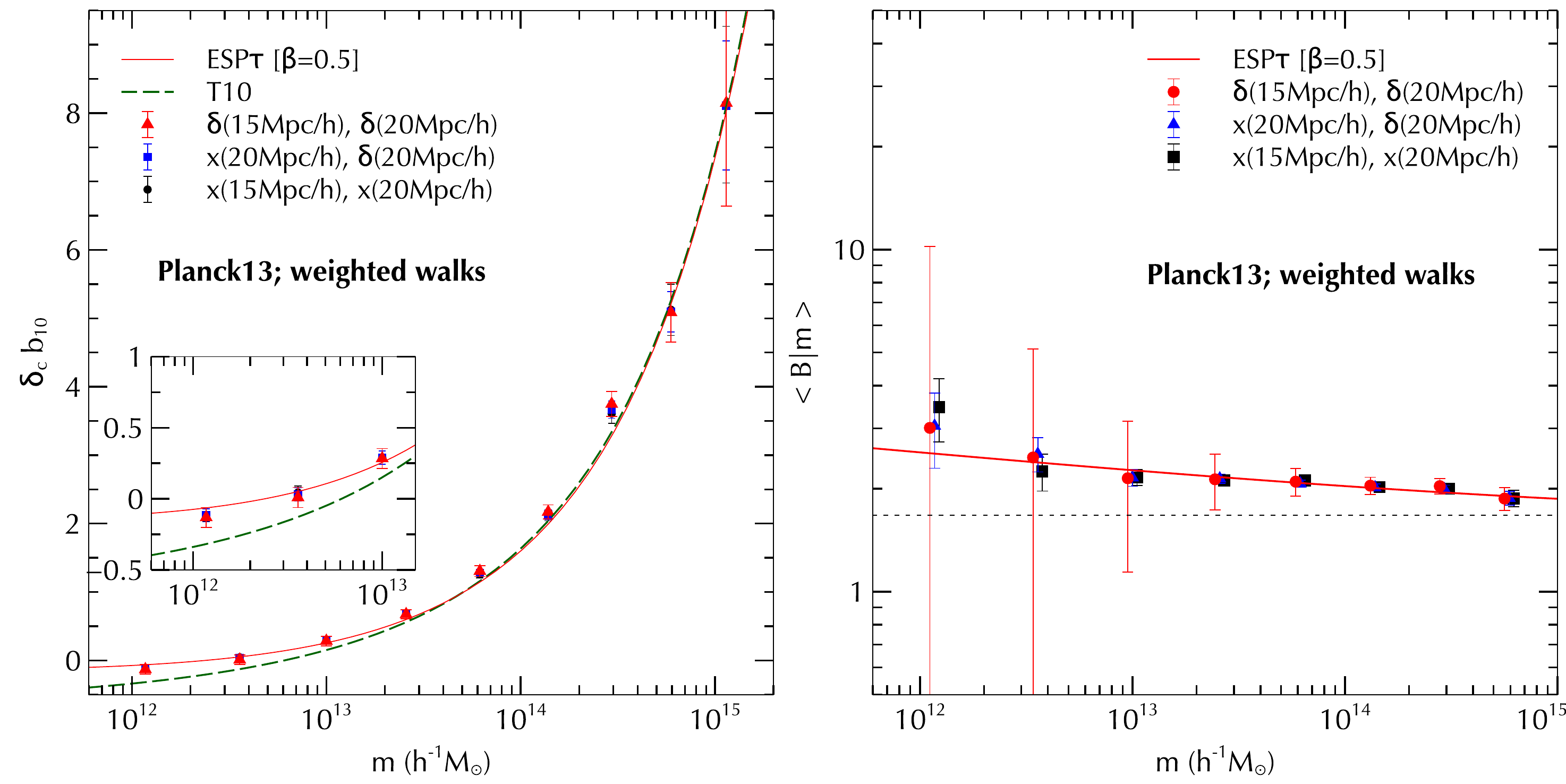}
\caption{Results using peak-weighted random walks. \emph{(Left panel:)} Peak-background split bias parameters $\delc b_{10}$. \emph{(Right panel:)} Model-independent recovery of mean barrier $\avg{B|m}$ from measurements of $b_{10}$ and $b_{11}$.}
\label{fig:biasfromwalks}
\end{figure*}

When computing bias, on the other hand, the weight is included in the estimators in \eqn{bn-estimator}. For instance, for $b_n$, in the $i^{\rm th}$ bin of smoothing scale,
\begin{align}
\hat{b}_n =  \frac{S_0^{n/2}}{S_\times^n} & \frac{\sum_{\alpha=1}^{N_i} w_{\rm pk}(x_{R_\alpha},R_\alpha) H_n(\delta_{0,\alpha}/\sqrt{S_0})}{\sum_{\alpha=1}^{N_i} w_{\rm pk}(x_{R_\alpha},R_\alpha)}\,.
\end{align}
Measurements such as these can be combined, as discussed in the main text, to provide numerical estimates of the scale independent coefficients $b_{nr}$ (see below).

Dealing with random walks also helps to assert the validity of our approximation of trading  $\tau$ with $q_5$, that we used to simplify the evaluation of the peak constraint. In the Monte Carlo runs with barrier as in \eqn{tau-barrier}, but with the true $q_5$ replacing $\tau$, we can actually weight by $F(x)\,\HT(x)$ as well as by the full $F(x,y,z)\chi_{\zeta}(x,y,z)$. Unfortunately, while the former can be applied to each single one of  the over a million walks we have first crossing for, only a few tens of thousands of these walks satisfy the latter constraint. Nevertheless, we find that, on the range of scales we expect our model to provide a good description of the haloes in the simulations, the difference between the full weighting of a $q_5$ barrier and the $\text{ESP}\tau$ barrier is negligible (see Figure~\ref{fig:fnufromwalks}). Departures start to become appreciable at $\nu<1$. 
For the rest of the Appendix we therefore focus on $F(x)$ weighing.

\begin{figure}
\centering
\includegraphics[width=0.45\textwidth]{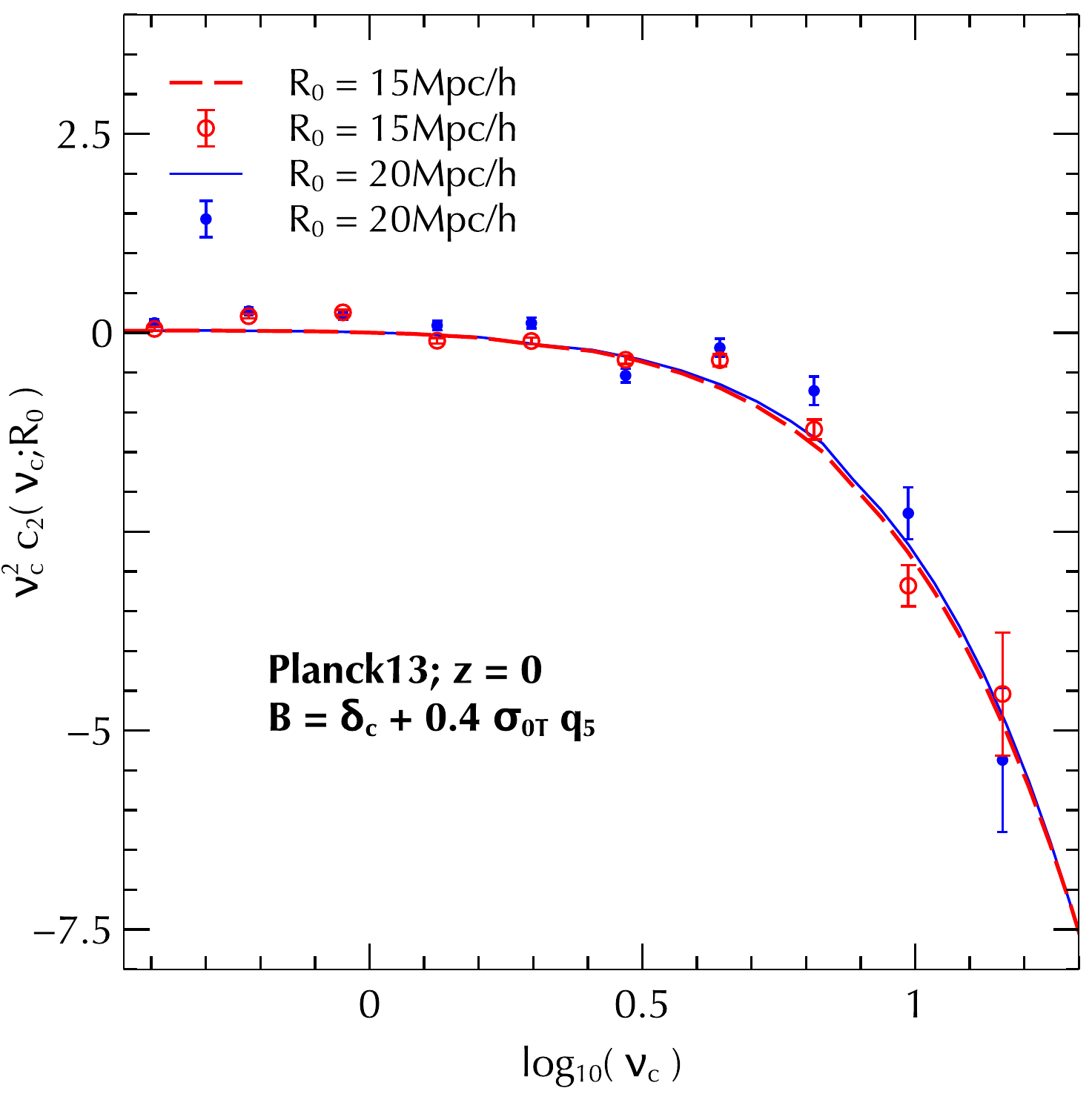}
\caption{Results using peak-weighted random walks. Nonlocal bias (points with errors) estimated at $R_0=15h^{-1}$Mpc (empty red circles) and $R_0=20h^{-1}$Mpc (filled blue circles). The corresponding smooth curves show the analytical results of \cetal\ with the extra peak weight. }
\label{fig:c2fromwalks}
\end{figure}

Figure~\ref{fig:biasfromwalks} compares measurements of $b_{10}$ and the model-independent estimate of the mean barrier $\avg{B|m}$ using the weighted walks described above, with the analytical results described in the main text. We see that the analytical model agrees extremely well with these measurements, including at small masses. The discrepancies seen between the model and the proto-halo based estimates of $b_{10}$ in the right panel of Figure~\ref{fig:b11-b10} are a failing of the basic underlying assumptions in the ESP approach, rather than of the various technical approximations made in the excursion set calculations. On the other hand, the good agreement between model and measurements in the right panel of Figure~\ref{fig:biasfromwalks} reassures us that the discrepancies seen in Figure~\ref{fig:meanbarrier} are in fact due to the biased nature of the estimate, as discussed in section~\ref{subsec:meanbarrier}. 

Given the noisy measurements of the shear bias coefficient $c_2$ in simulations seen in Figure~\ref{fig:c2}, we further tested our analytical predictions with the outcome of the Monte Carlo runs. The results are shown in Figure \ref{fig:c2fromwalks}. The ESP$\tau$ model provides a very good description of the measurements in the walks, with Figure~\ref{fig:c2fromwalks} also allowing us to show that the scale dependence of $c_2$ in the walks and in the analytic calculation (following \cetal, see also Appendix~\ref{app:nonlocalbias}) is negligible. Whether this result remains valid for protohaloes in an $N$-body simulation remains to be checked with more accurate measurements than the one we present in Figure~\ref{fig:c2}.

\label{lastpage}

\end{document}